\documentclass[aip,jap,preprint]{revtex4-1}  % for review and submission
\usepackage{graphicx}  % needed for figures
\usepackage{dcolumn}   % needed for some tables
\usepackage{bm}        % for math
\usepackage{subfigure}
\begin{document}

%=============================================INTRODUCTION===============================================

\title{Complex impedance, responsivity and noise of transition-edge sensors: analytical solutions for two- and three-block thermal models}

\author{I. J. Maasilta}
 \affiliation{Nanoscience Center, Department of Physics, P. O. Box 35, FIN-40014 University of Jyv\"askyl\"a, Finland}
 \email{maasilta@jyu.fi }

%=============================================INTRODUCTION===============================================

\begin{abstract}
The responsivity and noise of a voltage-biased superconducting transition-edge sensor depends strongly on the details of its thermal model, and the simplest theory for TES response assumes a single heat capacity connected to the heat bath. Here, analytical results are derived and discussed for the complex impedance, the responsivity and the noise of a transition-edge sensor, when the thermal model is not simple but consists of either two or three connected heat capacities. The implications of the differences of the models are discussed, as well. 
\end{abstract}

\keywords{transition-edge sensor,thermal model,noise,complex impedance}

\maketitle

\section{Introduction}
\label{intro}

Superconducting transition-edge sensors (TES) have become very popular as practical radiation detectors because of their high sensitivity and energy resolution, maturity of the fabrication technology required allowing wafer-scale processing, and versatility, so that they can be used for detection of both particles and photons from sub-millimeter frequencies to gamma-rays \cite{irwinbook}. The basic operational theory of a voltage-biased TES in electrothermal feedback was described long ago \cite{irwin95}, however, in the original form the simplest possible thermal circuit was assumed, namely, that the detector could be described as a single lumped heat capacity, connected to the heat bath by  a single thermal conductance. Sometimes this thermal model is fairly adequate in describing the detector response \cite{oneblock}, but some detector designs have been experimentally shown to behave in a more complex manner \cite{hoevers,zink,NASASaab,nasaIEEE,mikko,kimmo}. Therefore, theoretical modeling has been advanced in recent years to include more complex thermal circuits \cite{hoevers,galeazzi,zink,enectali,appel,takeinote,goldie,zhaothesis}.

One way to approach the problem is to generalize the problem fully to any number of heat capacity blocks and thermal conductances, and solve the obtained (large) linearized system of equations numerically \cite{effthesis,enectali}. This approach has the advantage that it is straightforward to move from simple models to more complex models within the same formalism, and for the most complex models this may be the only approach available. Nevertheless, for models of intermediate complexity, such as systems of two or three connected heat capacities, it is possible to calculate the detector properties analytically, as well. Analytical solutions have the advantage that they are easier to work with, can be used further for fitting of experimental data, and a lot of results can be calculated quickly. One can quickly compute, for example, how the detector response changes as a function of a certain parameter of the thermal model. 

Here, we present analytical solutions for the complex impedance, responsivity, and all the unavoidable components of the current noise spectral density of a TES detector, for all possible two-block thermal models and two simplest three-block models, in a compact formulation. We take advantage of the fact that we also derive the analytical equation for the complex impedance of the detector \cite{mather1,moseley,linde} of each model. Measurement of the complex impedance of a TES detector has been shown in recent years \cite{linde,vailla,zink,linde2} to be a very valuable tool for characterising TES detectors; in particular for the discussion here, it gives information about the thermal model. There is some previously published \cite{galeazzi,hoevers, appel,zhaothesis} and unpublished \cite{takeinote} analytical work on  certain two- and three-block models of bolometers. Those results describe some limits and special cases of this work. In Refs. \cite{takeinote,hoevers}, some two-block models were discussed, and an approximation was made about the steady state temperature of the blocks. In Refs. \cite{galeazzi,appel}, equations for the impedance, responsivity and the noise equivalent power were given for one two-block and three three-block models, but the non-Ohmic behavior of the bolometer \cite{mather,irwinbook} was not fully discussed. In Ref. \cite{zhaothesis}, two simple three-block models were discussed, but no simple analytical equations for the noise were given. The goal of the present work is thus to give an extensive set of equations  for the responsivity, current noise and impedance in a compact and usable form, which can easily be used to analyze real noise and impedance data of TES detectors \cite{mikko,kimmo}. We also give many example plots to show how various thermal parameters affect the detector properties. Detailed discussion on the noise equivalent power, energy resolution and other figures of merit are left for future publications.

\section{Two-block models}
\label{sec:1}

\begin{figure}
  \includegraphics[width=\textwidth]{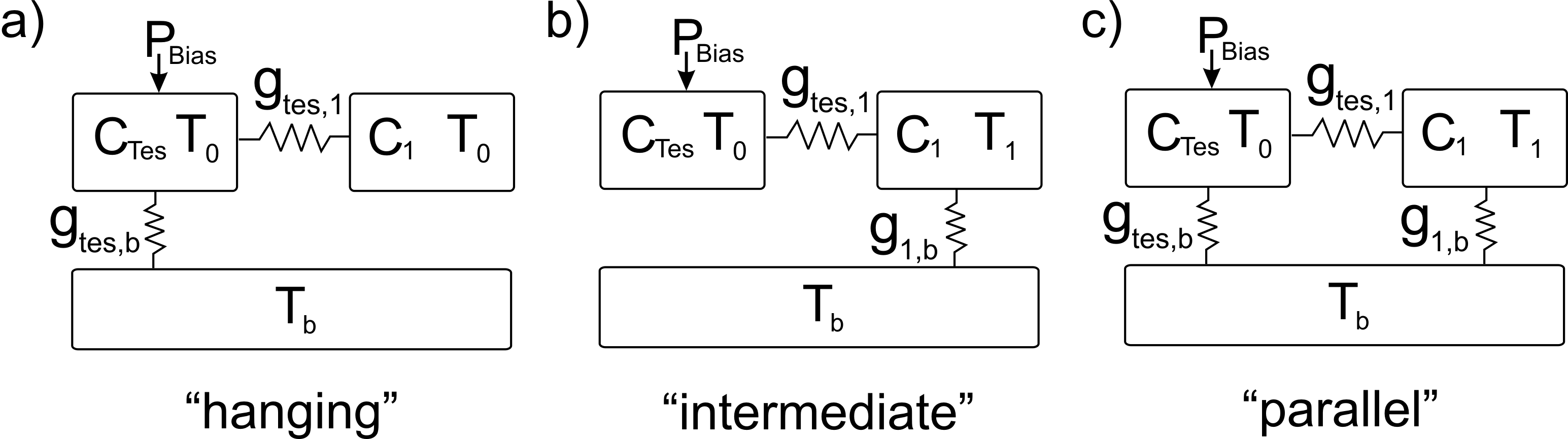}
\caption{All possible two-block models studied. (a) Hanging model, (b) intermediate model, (c) parallel model.}
\label{twoblocks}       % Give a unique label
\end{figure}

We discuss here all three possible two-block cases, shown in Fig \ref{twoblocks}. One heat capacity block is always the TES film, where the Joule power is dissipated, whereas the second block describes an additional thermal body, which could represent for example the insulating membrane, the absorber etc. \cite{mikko,kimmo,nasaIEEE,goldie}. The formalism is kept general so that we do not need to decide on the physical picture {\em a priori}. 

The first case, shown in Fig. \ref{twoblocks} (a), is where the extra heat capacity $C_1$ is connected to the TES heat capacity $C_{tes}$, but nowhere else. This type of model has sometimes been called the hanging or dangling heat capacity model. In the second case, shown in Fig. \ref{twoblocks} (b), the extra heat capacity lies in between the TES and the heat bath. The main difference to the first model is that now the steady state temperature of the extra block is not equal to the TES temperature, unlike in the first case of Fig. \ref{twoblocks} (a).  We call this second model the intermediate model. 
Finally, in the last model in Fig. \ref{twoblocks} (c) (numerically studied in Ref. \cite{goldie}), the extra block is connected to both the TES and the heat bath  as in case (b), but an additional parallel heat path exists between the TES and the bath. We call this the parallel model.

\subsection{Equations to be solved}
\label{sec:3}

The starting point for all calculations are the differential equations that describe the heat balance in the thermal circuits and the electrical equation of the simplest Thevenin-equivalent input circuit of the TES, consisting here of an equivalent voltage $V_b$, an equivalent resistance $R_L$, and an inductance $L$ in series.  For a simple one-block thermal circuit, there is only one thermal equation, so that one needs to solve  a coupled set of only two differential equations, as reviewed in Ref.  \cite{irwinbook}. They are coupled because the resistance of the TES detector $R(T,I)$ depends on temperature and current, and because the Joule heating power $P_J$ in the TES depends on the electrical parameters. It is this coupling that produces the negative electrothermal feedback in the response of a voltage biased superconducting detector \cite{irwin95,irwinbook}. For a two-block thermal model, one additional differential equation is added to represent the extra thermal body. The exact form of the thermal equations naturally depends on which model in Fig. \ref{twoblocks} one is analyzing. Here, we only show the equations for case (a), the hanging model,  other cases follow analogously. The set of equations, excluding the noise terms for a moment, reads
\begin{eqnarray}
C_{tes}\frac{dT_{tes}}{dt} &=& P_{J}-A\left (T_{tes}^n-T_1^n \right )-B\left (T_{tes}^m-T_b^m \right )+\delta P, \nonumber\\
C_{1}\frac{dT_{1}}{dt} &=& A\left (T_{tes}^n-T_1^n \right ), \nonumber\\
L \frac{dI}{dt} &=& V_b-IR_L-IR(T_{tes},I),
\label{diffeqs}
\end{eqnarray}

where $T_{tes}$ and $T_{1}$ are the instantaneous temperatures of the TES and the extra heat capacity, respectively, and the heat bath is at a temperature $T_b$. $A$ and $B$ are constants describing the strength of the thermal links between the TES and $C_1$, and TES and heat bath, respectively, and $n$ and $m$ are temperature exponents, which depend on the physical nature of the thermal links \cite{irwinbook}. $\delta P$, on the other hand, is some time-dependent power input to the TES. Using these equations, the full time-dependent response (temperature and current) could, in principle, be solved. However, the power flows are non-linear functions of temperature, and the TES resistance is a non-linear function of both $T$ and $I$, so that an analytical solution for the system is not tractable in the general (large signal) case. In this work, however, we are only interested in the impedance, noise and responsivity, all of which are traditionally defined only in the small-signal limit, where the non-linearities can be ignored. The linearization of Eqs. \ref{diffeqs} is done by the usual Taylor expansions around steady state values \cite{irwinbook} (denoted by subscript $0$): 
\begin{eqnarray}
IR(T_{tes},I) &\approx& I_{0}R_0+\alpha I_{0}(R_0/T_{tes,0}) \Delta T_{tes}+(1+\beta)R_{0}\Delta I, \nonumber\\ 
P_{J} &\approx& P_{0}+[2I_{0}R_{0}+\beta (P_{0}/I_{0})]\Delta I+\alpha (P_{0}/T_{tes,0}) \Delta T_{tes}, \nonumber\\  
A(T_{tes}^n-T_1^n) &\approx& A(T_{tes,0}^n-T_{1,0}^n)+nAT_{tes,0}^{n-1} \Delta T_{tes}-nAT_{1,0}^{n-1} \Delta T_{1}, \nonumber\\ 
B(T_{tes}^m-T_b^m) &\approx& B(T_{tes,0}^m-T_{b}^m)+mBT_{tes,0}^{m-1} \Delta T_{tes},
\end{eqnarray}
where $\Delta T_{tes}= T_{tes}-T_{tes,0}$, $\Delta I=I-I_0$, $\Delta T_{1}= T_{1}-T_{1,0}$, and $\alpha=(T_{tes,0}/R_0)(\partial R/\partial T)|_{I_{0}}$ and $\beta=(I_{0}/R_0)(\partial R/\partial I)|_{T_{tes,0}}$ are the dimensionless (logarithmic) transition sensitivity parameters. Substituting the above expansions into Eqs. \ref{diffeqs} lead to the linearized set of equations
\begin{eqnarray}
C_{tes}\frac{d\Delta T_{tes}}{dt} &=& \left ( 2I_{0}R_{0}+\beta \frac{P_{0}}{I_{0}} \right )\Delta I + \left ( \alpha \frac{P_{0}}{T_{tes,0}}-g_{tes,1}-g_{tes,b} \right ) \Delta T_{tes}+ g_{tes,1}\Delta T_{1} +\delta P, \nonumber\\
C_{1}\frac{d\Delta T_{1}}{dt} &=&  g_{tes,1} \left ( \Delta T_{tes}-\Delta T_{1} \right ), \nonumber\\
L \frac{d \Delta I}{dt} &=& \delta V_b-\Delta IR_L-\alpha \frac{I_{0}R_{0}}{T_{tes,0}} \Delta T_{tes}-(1+\beta)R_{0}\Delta I,
\label{lindiffeqs}
\end{eqnarray}

where we have defined the differential thermal conductances $g_{tes,1}=nAT_{tes,0}^{n-1}$ and $g_{tes,b}=mBT_{tes,0}^{m-1}$, and used the fact that for the hanging model $T_{1,0}=T_{tes,0}$. From now on, we also simplify the notation as $T_{0} \equiv T_{tes,0}$. Also, after linearization, one can ignore the temperature dependence of the heat capacities and consider them to be constants [$C_{tes}=C_{tes}(T_{0})$, $C_{1}=C_{1}(T_{0})$ ], as the corrections are in second order \cite{effthesis}.

Eqs. \ref{lindiffeqs} have been analyzed in time-domain in Ref \cite{bennett}, with focus on pulse response and electrothermal stability analysis. Here, we proceed to focus on the frequency domain (also discussed in Ref. \cite{takeinote} for this model), which is natural for impedance and noise analysis. Fourier transforming Eqs. \ref{lindiffeqs} and simplifying notation, one gets 
\begin{eqnarray}
i\omega T_{\omega,tes} &=& \frac{I_{0}R_{0}(2+\beta)}{C_{tes}}I_{\omega}- \frac{1}{\tau_I}T_{\omega,tes}+\frac{1}{\tau_{tes,1}}T_{\omega,1}+\frac{1}{C_{tes}}P_{\omega}, \nonumber\\
i\omega T_{\omega,1} &=& \frac{1}{\tau_{1}}\left ( T_{\omega,tes}-T_{\omega,1} \right ) \nonumber\\
i\omega L I_{\omega} &=& V_{\omega}-\left [ R_{L}+R_{0}(1+\beta) \right ] I_{\omega}-\frac{{\cal L}_{H}(g_{tes,1}+g_{tes,b})}{I_{0}} T_{\omega,tes},
\label{fouriereqs}
\end{eqnarray}
where we have denoted the Fourier amplitudes of all variables as $X_{\omega}$, have defined $\tau_{tes,1}= C_{tes}/g_{tes,1}$, and have defined two other important time constants and a dimensionless quantity ${\cal L}_{H}$ that appears  in the place of the loop gain of the simple model as
\begin{eqnarray}
\tau_{I} &=& \frac{C_{tes}}{(g_{tes,1}+g_{tes,b})(1-{\cal L}_{H})}, \nonumber\\
\tau_{1} &=& \frac{C_{1}}{g_{tes,1}}, \nonumber\\
{\cal L}_{H} &=& \frac{P_{0}\alpha}{(g_{tes,1}+g_{tes,b})T_{0}}.
\label{symbolsH}
\end{eqnarray}

Note that the definition of this "effective loop gain" ${\cal L}_{H}$ here is different from the standard one-block model one \cite{irwinbook} ${\cal L} = P_{0}\alpha/(g_{tes,b}T_{0})$, and therefore ${\cal L}_{H} = g_{tes,b}/(g_{tes,1}+g_{tes,b}){\cal L}$. In fact, if one calculates the true zero frequency, perfect voltage bias loop gain starting from the definitions \cite{mccammonbook}, one arrives at ${\cal L}$ even for this hanging model. This is intuitive, as the coupling to bath is the same. However, the new definition does have the benefit that for each different model (hanging, intermediate or parallel), the equations \ref{fouriereqs} have the same mathematical form, only the definition of the effective loop gain and the time constants change, as will be seen later.  It is, of course, still possible to work with other, less compact notations, as was done for example in Refs. \cite{bennett,takeinote,zhaothesis}.

\subsection{Complex impedance}

\subsubsection{Hanging model}

\begin{figure}[p]
  \includegraphics[width=0.9\textwidth]{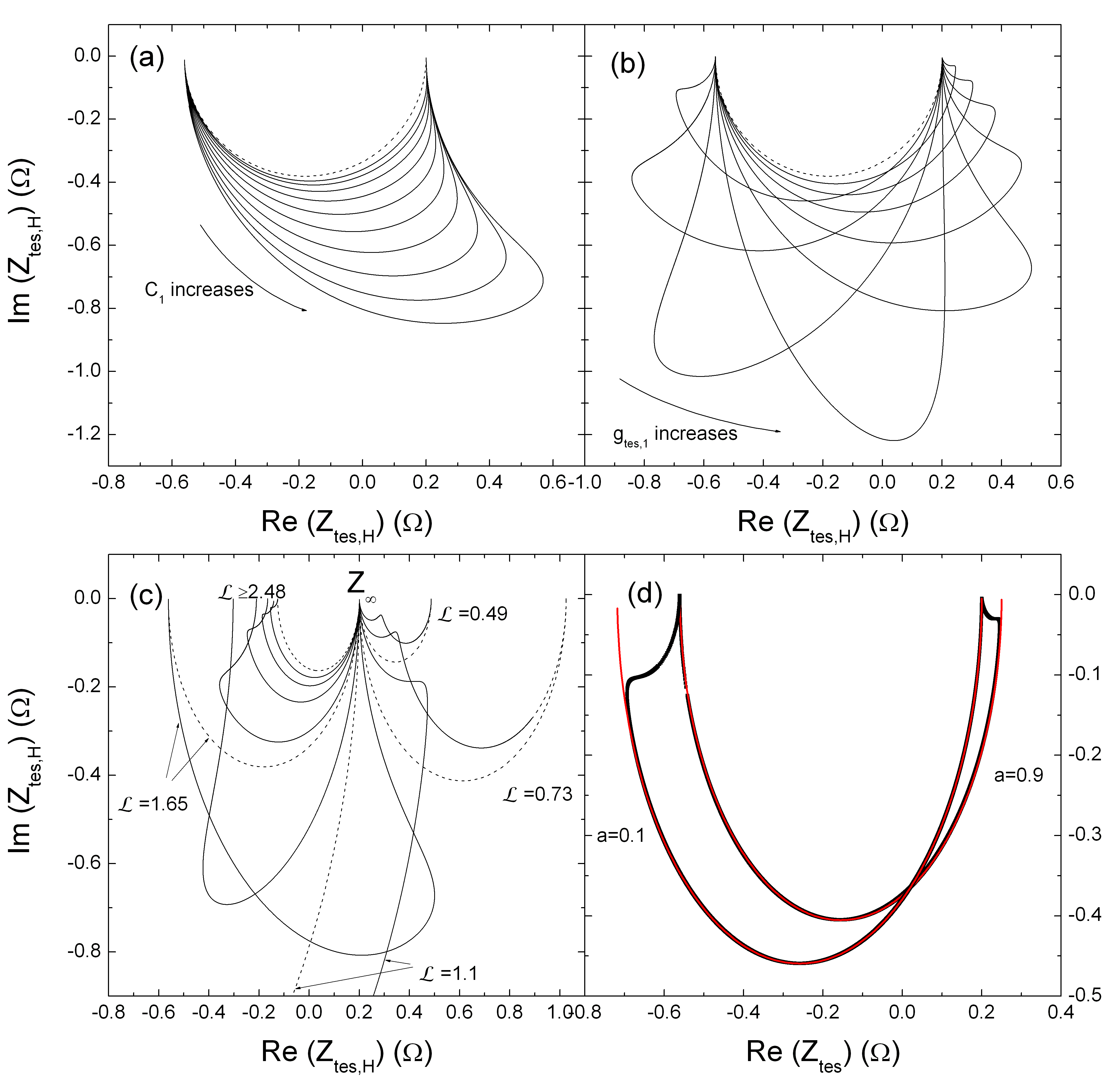}
\caption{(Color online) Complex impedance of a two-block hanging model, with varying parameters. We have used $R_{0}=0.1 {\mathrm \Omega}$ and $\beta=1$ in all plots, and frequencies run typically between $\omega\tau_{tes} = 0.01 .. 100$, where $\tau_{tes}=C_{tes}/g_{tes,b}$. (a) $Z_{tes,H}$ as a function of $C_{1}$, with $C_{1}/C_{tes}= 0.33, 0.49, 0.73, 1.1, 1.65, 2.48, 3.71, 5.57, 8.35, 12.5$, ${\cal L}=1.65$ and $a=g_{tes,1}/(g_{tes,1}+g_{tes,b})=0.5$. Increasing $C_{1}$ corresponds to a larger deviation from the one-block model, shown as dashed line. (b) $Z_{tes,H}$ as a function of $a$ ( $g_{tes,1}$), with $a=0.1, 0.2, 0.3, 0.4, 0.5, 0.6, 0.7, 0.8, 0.9$, ${\cal L}=1.65$ and $C_{1}/C_{tes}=10$. Increasing $a$ corresponds to the bulge feature moving from the left, low-frequency side to the right, high-frequency side. Dashed line is the one-block model limit ($a \rightarrow 0$). (c) $Z_{tes,H}$ as a function of ${\cal L} = P_{0}\alpha/(g_{tes,b}T_{0})$, with ${\cal L}= 0.49, 0.73, 1.1, 1.65, 2.48, 3.71, 5.57, 8.35, 12.5$, $C_{1}/C_{tes}=10$ and $a=0.5$. Dashed lines show the corresponding one-block results. (d) Comparison of two $Z_{tes,H}$ impedance curves (points) ($C_{1}/C_{tes}=10$, ${\cal L}= 1.65$ and either $a=0.1$ or $a=0.9$) with one-block model fits (red lines). For $a=0.9$, the one-block fitted parameters are $\beta=1.5$, ${\cal L}=1.76$ and $C= 12 C_{tes}$, whereas for $a=0.1$, we get fitted values ${\cal L}=1.485$ and $C= 0.9 C_{tes}$, with $\beta=1$ kept fixed.}
\label{plotsH}       % Give a unique label
\end{figure}

The first quantity we want to derive from Eqs. \ref{fouriereqs} for the hanging model is the frequency dependent complex impedance of the TES, $Z_{tes,H}$. As the full circuit impedance $Z_{circ}$ is calculated by $Z_{circ}=V_{\omega}/I_{\omega}$, we can subtract from it the known impedance of the circuit outside of the TES to define $Z_{tes}=Z_{circ}-R_{L}-i\omega L$ for the simplest bias circuit case, or $Z_{tes}=Z_{circ}-Z_{bias}$ in general. We can set $P_{\omega}=0$, and obtain 
\begin{equation}
Z_{tes,H}=R_0(1+\beta)+\frac{{\cal L}_{H}}{1-{\cal L}_{H}}R_0(2+\beta) \left / \left [1+i\omega\tau_{I}-\frac{g_{tes,1}}{(g_{tes,1}+g_{tes,b})(1-{\cal L}_{H})}\frac{1}{1+i\omega\tau_{1}}
 \right ] \right.,
\label{Zhang}
\end{equation}
with definitions of ${\cal L}_{H}$, $\tau_{I}$ and $\tau_{1}$ given in Eqs. \ref{symbolsH}.
This can be compared with the result for the simple one-block model \cite{irwinbook} 
\begin{equation}
Z_{tes}=R_0(1+\beta)+\frac{{\cal L}}{1-{\cal L}}R_0(2+\beta) \frac{1}{1+i\omega\tau_{I}}, 
\label{Zsimple}
\end{equation}
where $\tau_{I}=C_{tes}/[g_{tes,b}(1-{\cal L})]$ for the one-block model. We see that due to the second heat capacity, a new frequency-dependent term with time constant $\tau_{1}$ appears in the denominator. Thus, the effect of $C_{1}$ is non-linear. The strength of the extra term is not only set by the value of $C_{1}$, but also by how large the thermal conductance $g_{tes,1}$ is relative to $g_{tes,b}$. However, as the effective loop gain ${\cal L}_{H}$ also depends on the ratio of the two thermal conductances, it turns out that $Z_{tes,H}$ deviates maximally in the complex plane from the simple model for some value $a=g_{tes,1}/(g_{tes,1}+g_{tes,b}) < 1$, depending on the value of loop gain. In the limits $a \rightarrow 0$ and $a \rightarrow 1$, $Z_{tes,H}$ approaches the simple one-block model result, with a heat capacity $C_{tes}$ and $C_{tes}+C_{1}$, respectively. We show some example plots of the effects of $C_{1}$ and $g_{tes,1}$ in Fig. \ref{plotsH}. We see that increasing $C_{1}$ distorts $Z_{tes,H}$ more from the simple-model half-circle [Fig. \ref{plotsH} (a)], but keeps the direction of the extra "bulge" constant in the complex plane. Changing $g_{tes,1}$, on the other hand, changes both the location and the size of the "bulge" feature.  The loop gain also has a direct and quite complex effect, as shown in Fig. \ref{plotsH} (c). In the large loop gain limit, the two-block hanging model approaches the simple model, but for typical values of  ${\cal L}$ the effect of the extra term is strong, and, interestingly, remains even in the limit ${\cal L} \rightarrow 0$. For $0 < {\cal L} < 1$, the effect of the two-block model is opposite compared to ${\cal L} > 1$, in that the circular shape of the simple model is distorted inward. 

In Fig. \ref{plotsH} (d) we show two examples of a potential pitfall in analyzing impedance data. In the limits where $g_{tes,1}$ is either clearly smaller or clearly larger than $g_{tes,b}$, the shape of $Z_{tes,H}$ is only distorted at the low-frequency or high-frequency end, respectively. In that case, an incomplete frequency range of the data could lead to a misinterpretation of a two-block model as a one-block model. For example, in Fig. \ref{plotsH} (d) we show a plot of a two-block model with $a=0.9$, $\beta=1$, ${\cal L}=1.65$, and $C_{1}/C_{tes}=10$, which can be fitted with a one-block model up to a frequency $\omega\tau_{tes} \sim 2$, but with misidentified $\beta=1.5$, ${\cal L}=1.76$ and $C= 12 C_{tes}$. The same way, the two-block impedance with $a=0.1$ (other parameters the same) can be fitted with a one-block model with misidentified ${\cal L}=1.485$ and $C= 0.9 C_{tes}$, down to a frequency  $\omega\tau_{tes} \sim 0.05$. The high-frequency distortions can get especially tricky, as measurements always have an electrical high-frequency cut-off somewhere.   

\subsubsection{Intermediate model}

Turning to the second, intermediate model (Fig. \ref{twoblocks} (b)), we can follow similar derivation as for the hanging model. Omitting details, we find the same mathematical form for $Z_{tes}$, but with different definitions of the effective loop gain and time constants: 
\begin{equation}
Z_{tes,IM}=R_0(1+\beta)+\frac{{\cal L}_{IM}}{1-{\cal L}_{IM}}R_0(2+\beta) \left / \left [1+i\omega\tau_{I}-\frac{g_{tes,1}(T_{1})}{[g_{tes,1}(T_{1})+g_{1,b}](1-{\cal L}_{IM})}\frac{1}{1+i\omega\tau_{1}} \right ] \right.,
\label{ZITF}
\end{equation}
where 
\begin{eqnarray}
\tau_{I} &=& \frac{C_{tes}}{g_{tes,1}(T_{0})(1-{\cal L}_{IM})} \nonumber\\
\tau_{1} &=& \frac{C_{1}}{g_{tes,1}(T_{1})+g_{1,b}} \nonumber\\
{\cal L}_{IM} &=& \frac{P_{0}\alpha}{g_{tes,1}(T_{0})T_{0}}.
\end{eqnarray}

Here, we must also explicitly define at what temperature $g_{tes,1}$ is evaluated, as the two blocks have different steady-state temperatures in this case ($T_0$ for $C_{tes}$ and $T_{1}$ for $C_{1}$), and therefore there are two different values for $g_{tes,1}$ that come into play: $g_{tes,1}(T_{1})$ and $g_{tes,1}(T_{0})$. $g_{1,b}$ is always evaluated at $T_{1}$, and is therefore not explicitly written in the above equations for simplification. $g_{1,b}(T_{b})$ will only come into play when calculating noise (Sect. \ref{noisesection}).

Eq. \ref{ZITF} thus shows that the complex impedance of the intermediate model behaves qualitatively exactly like the hanging model one, so that similar plots to Fig. \ref{plotsH} can be generated, but with different parameter values. It is therefore very hard, if not impossible, to distinguish between the hanging and intermediate models based on fitting impedance data alone, as was pointed out in Refs. \cite{takeinote,lindeIEEE} already.  

In the intermediate model, it is clear that the true loop gain is really different from the simple model, unlike in the hanging case, as the extra block affects the DC response. One way to find the loop gain is to look at the zero frequency limit of Eq. \ref{ZITF}. Doing that, one finds ${\cal L}=P_{0}\alpha/(g_{eff}T_{0})$, where 
\begin{equation}
g_{eff}=\frac{ g_{tes,1}(T_{0})g_{1,b}}{g_{tes,1}(T_{1})+g_{1,b}} 
\end{equation}
is an effective differential thermal conductance, which is a series combination of the two individual thermal conductances. Notice, though, how $g_{tes,1}$ has to be evaluated at different temperatures in the nominator and in the denominator. The added complication for analysis is that if $T_{1}$ is not directly measurable (typical case), then the DC I-V measurements alone cannot fix $T_0$ or the parameters $A$, $B$, $n$ and $m$, as there are too many unknowns in the general case. Only if one can set $n=m$ based on physical assumptions (same thermal conduction mechanism), can one determine both $T_0$ and $g_{eff}$ from the I-V data. $T_{1}$ naturally depends on the relative strength of the two thermal conductances such that if $n=m$,

\begin{equation}
T_{1}=\left ( \frac{g_{tes,1}(T_{1})}{g_{tes,1}(T_{1})+g_{1,b}} T_0^n+ \frac{g_{1,b}}{g_{tes,1}(T_{1})+g_{1,b}} T_{b}^n \right )^{1/n}.
\label{T1eq}
\end{equation}

%Moreover, there is still a strong correlation between the two free parameters $T_{1}$ and $C_{1}$, as $\tau_{1} \sim C_{1}/T_{1}^{n-1}$ (both parameters change the impedance the same way only through $\tau_{1}$), leading to ambiguity in fixing their values based on $Z_{tes}$ data alone. Lowering $T_{1}$ is thus equivalent to increasing $C_{1}$, both leading to enhancement of the deviation from one-block results. 

\subsubsection{Parallel model}

Finally, we discuss the third and final two-block model, where both blocks have conduction channels to the bath, the parallel model [Fig. \ref{twoblocks} (c)]. Again, it is straightforward to derive equations analogous to Eqs. \ref{fouriereqs} and to solve for $Z_{tes}$, yielding once again the same form
\begin{eqnarray}
Z_{tes,P}&=&R_0(1+\beta)+\frac{{\cal L}_{P}}{1-{\cal L}_{P}}R_0(2+\beta) /  \nonumber\\ 
& & \left [1+i\omega\tau_{I}-\frac{g_{tes,1}(T_{0})g_{tes,1}(T_{1})}{[g_{tes,1}(T_{0})+g_{tes,b}][g_{tes,1}(T_{1})+g_{1,b}](1-{\cal L}_{P})}\frac{1}{1+i\omega\tau_{1}} \right ],
\label{parallel}
\end{eqnarray} 
with definitions
\begin{eqnarray}
\tau_{I} &=& \frac{C_{tes}}{(g_{tes,1}(T_{0})+g_{tes,b})(1-{\cal L}_{P})} \nonumber\\
\tau_{1} &=& \frac{C_{1}}{g_{tes,1}(T_{1})+g_{1,b}} \nonumber\\
{\cal L}_{P} &=& \frac{P_{0}\alpha}{(g_{tes,1}(T_{0})+g_{tes,b})T_{0}}.
\label{parasymbols}
\end{eqnarray}
 
Comparing with the two previous models, the amplitude factor for the extra term is naturally more complex (depending on all $g$s), but $\tau_{I}$ and ${\cal L}_{P}$ are the same as for the hanging model, whereas $\tau_{1}$ is the same as for the intermediate model. Again, similar plots could be produced as for the two previous models, and thus distinguishing the parallel model from the hanging and intermediate models by fitting impedance data alone is quite hopeless. In the parallel model, the true loop gain is still ${\cal L}=P_{0}\alpha/(g_{eff}T_{0})$, but now the effective conductance is slightly more complex, as it is a combination of parallel and series components:
\begin{equation}
g_{eff}=\frac{g_{tes,1}(T_{0})g_{1,b}}{g_{tes,1}(T_{1})+g_{1,b}}+g_{tes,b}. 
\end{equation}
Naturally, similar analysis issues exist for this model as was discussed for the intermediate model.    

\subsection{Small-signal responsivity}

Next, we turn from complex impedance to small-signal current responsivity, which describes the frequency-dependent current response of the device to power input to the TES. Naturally, one can define responsivities for power inputs to the other heat capacity blocks as well, but here we concentrate only in the direct responsivity, as it is important for the noise analysis discussed in the next section. The responsivity for each variant of the two-block models can be calculated again from equations \ref{fouriereqs}, by keeping the power term $P_{\omega}$ and by dropping the voltage bias modulation term $V_{\omega}$. Then, defining the responsivity as $s_I(\omega)=I_{\omega}/P_{\omega}$, one gets a result that looks a lot like $Z_{tes}$, as we start from almost the same equations. Therefore, for all models, we can write the responsivity as a function of $Z_{tes}$ and the total circuit impedance $Z_{circ}=Z_{tes}+Z_{bias}$ in a compact form as
\begin{equation}
s_I(\omega)=-\frac{1}{Z_{circ}I_0}\frac{Z_{tes}-R_0(1+\beta)}{R_0(2+\beta)},
\label{respo}
\end{equation}  
where for the simplest bias circuit we have  $Z_{circ}=Z_{tes}+R_{L}+i\omega L$. Eq. \ref{respo} above has the advantage that if the TES impedance is measured, as is commonly done, the responsivity can be immediately calculated from it. 

\begin{figure}[ht]
  \includegraphics[width=\textwidth]{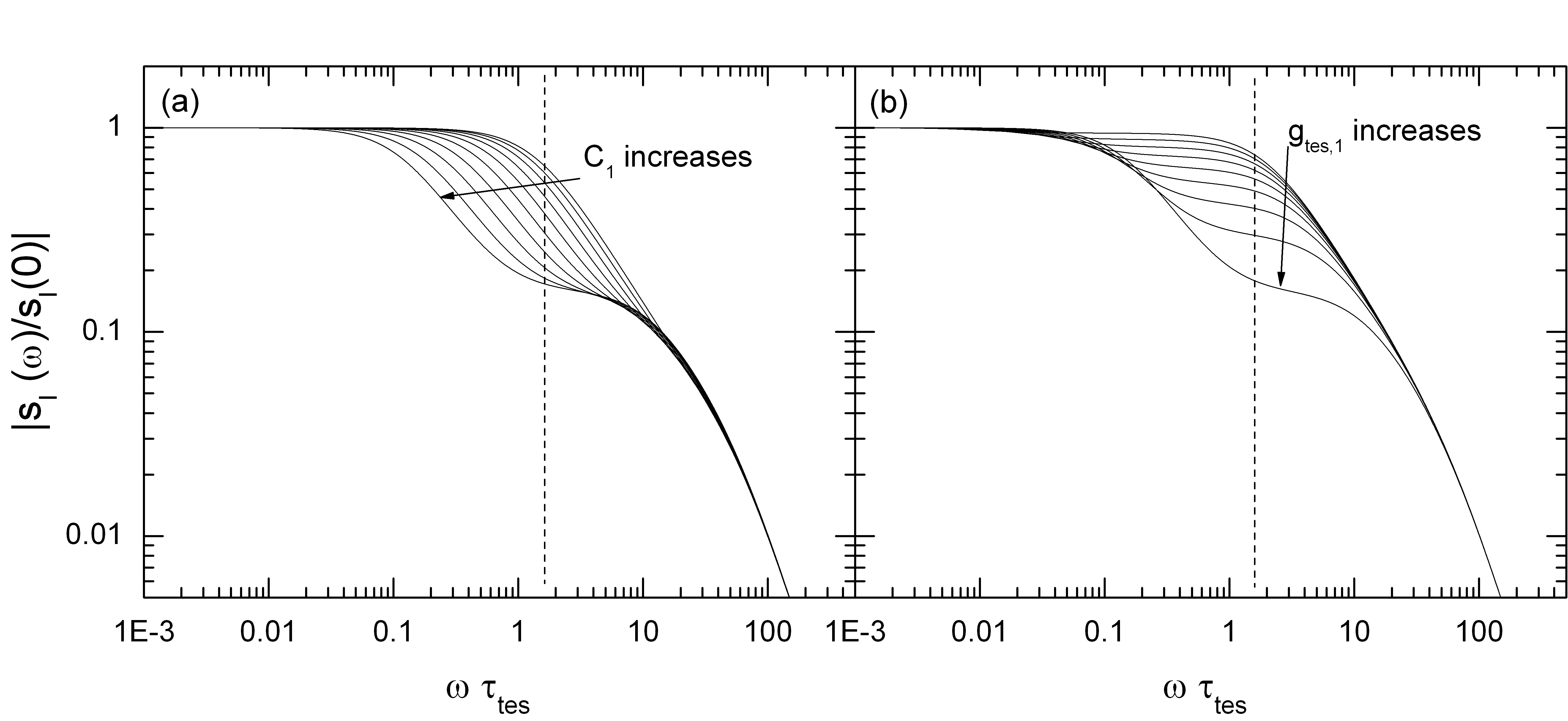}
\caption{ Responsivity of a two-block hanging model, with varying parameters. In both plots, we have used $R_{0}=0.1 \mathrm{\Omega}$, $\beta=1$, and $\tau_{el}=L/[R_L+R_0(1+\beta)]=0.015\tau_{tes}$, where $\tau_{tes}=C_{tes}/g_{tes,b}$.  (a) $|s_I(\omega)|$ as a function of $C_{1}$, with $C_{1}/C_{tes}= 0.33, 0.49, 0.73, 1.1, 1.65, 2.48, 3.71, 5.57, 8.35, 12.5$, ${\cal L}=1.65$ and $a=g_{tes,1}/(g_{tes,1}+g_{tes,b})=0.9$. Increasing $C_{1}$ corresponds to a shift of the partial thermal cut-off to lower frequencies, and the development of a "knee" in the intermediate frequency range. (b) $|s_I(\omega)|$  as a function of $a$ (or $g_{tes,1}$), with $a=0.1, 0.2, 0.3, 0.4, 0.5, 0.6, 0.7, 0.8, 0.9$, ${\cal L}=1.65$ and $C_{1}/C_{tes}=10$. Increasing $a$ corresponds to the decrease of responsivity in the intermediate frequency range. Dashed line shows the simple model effective thermal time constant.}
\label{SplotH}       % Give a unique label
\end{figure}

The effect of the loop gain on the current responsivity is well known, with higher loop gain increasing the responsivity at frequencies below the effective thermal time constant (where the responsivity starts to roll-off), and moving the time constant to higher frequencies. This takes place regardless of the complexity of the thermal model. On the other hand, the effect of changing $C_{1}$ and $g_{tes,1}$ is not so self-evident. In Fig. \ref{SplotH}, we plot how $C_{1}$ and $g_{tes,1}$  in the two-block thermal circuits affect the responsivity, using the hanging model as an example (other models behave in a similar manner). We plot only the magnitude $|s_I(\omega)|$, as the phase is irrelevant for noise considerations, which is the focus in this work. We see from the plots that for the usual case where the electrical cut-off frequency is above the thermal cut-off frequency (low enough inductance $L$), the effect of the extra thermal block is to reduce the responsivity in the mid- frequency range starting from  $\tau_{1}^{-1}$. Increasing $C_{1}$ moves the first partial cut-off set by $\tau_{1}^{-1}$ to lower frequencies [Fig. \ref{SplotH} (a)], whereas increasing $g_{tes,1}$ makes the intermediate frequency suppression stronger [Fig. \ref{SplotH} (b)].  

\subsection{Thermodynamic noise}
\label{noisesection}

In TES detectors, there are several possible mechanisms for noise. Some of them are directly associated with superconductivity such as phase-slip shot noise \cite{psn1,psn2},  fluctuation superconductivity noise \cite{fsn,seidel} and vortex motion noise \cite{vortex}.  In all these mechanisms, there are external parameters such as the material quality of the superconducting film or the magnetic field that affect the noise, and thus these superconductivity related noise sources can, in principle, be controlled to some extent. In addition, there is currently no conclusive evidence for a significant role of the superconducting noise sources being dominant in TES detectors. In this paper we therefore only concentrate on  the two unavoidable noise sources in TES detectors: (i) thermodynamic fluctuations of heat within the thermal circuit (thermal fluctuation noise, TFN) \cite{mather1}, and  (ii) electrical fluctuations of the resistive circuit elements (Johnson noise) \cite{mather1,johnson,nyquist}. 

The thermal fluctuation noise and Johnson noise of a simple one-block bolometer were discussed long ago \cite{mather1}, and their effect on a TES detector in negative electrothermal feedback has been reviewed thoroughly in Ref. \cite{irwinbook}. TFN and Johnson noise properties of some two- and three-body models have also been discussed in Refs. \cite{hoevers,zink,galeazzi,takeinote,zhaothesis}, but none of those Refs. give very simple compact analytical formulas for noise, nor do they discuss all the models covered in this work. In this section, compact analytical results for the current noise are given for all three variations of the two block models (and in Sect. \ref{sec:2} for two three-block models), using the already derived results for the complex impedance and responsivity.   
 
\subsubsection{Hanging model}
\label{hangsect}

\paragraph{Thermal fluctuation noise}

For each thermal conductance in the model, there are power fluctuations on top of the steady state power that is conducted. These fluctuations can be taken into account in a straightforward manner by adding appropriate fluctuation power terms $\delta P_i$ into the constituting thermal equations in Eqs. \ref{diffeqs}, which for the hanging model then become (ignoring external power input)
\begin{eqnarray}
C_{tes}\frac{dT_{tes}}{dt} &=& P_{J}-A\left (T_{tes}^n-T_1^n \right )-\delta P_{tes,1} -B\left (T_{tes}^m-T_b^m \right )-\delta P_{tes,b} \nonumber\\
C_{1}\frac{dT_{1}}{dt} &=& A\left (T_{tes}^n-T_1^n \right )+ \delta P_{tes,1}.
\label{diffeqs2}
\end{eqnarray}
Note how the sign has to be different for the $\delta P_{tes,1}$ terms in the two equations, because an increase in outflowing power from $i \rightarrow j$ is equivalent with a decrease of the outflowing power from $j \rightarrow i$. Linearization and transformation into frequency domain, as before, thus lead to equations that resemble the equations for complex impedance and responsivity. After some algebra, one can write surprisingly simple results (disregarding correlations between the the two fluctuating sources) for the two TFN current noise spectral densities $S_{I,i}(\omega)=|I_{\omega}|_{i}^2$ associated with the two thermal links $g_i$: 
\begin{eqnarray}
|I_{\omega}|^2_{tes,b}&=&\frac{P_{tes,b}^2}{|Z_{circ}|^2I_0^2}\frac{|Z_{tes}-R_0(1+\beta)|^2}{R_0^2(2+\beta)^2}= P_{tes,b}^2 |s_I(\omega)|^2, \nonumber\\
|I_{\omega}|^2_{tes,1}&=&\frac{P_{tes,1}^2}{|Z_{circ}|^2I_0^2}\frac{|Z_{tes}-R_0(1+\beta)|^2}{R_0^2(2+\beta)^2}\frac{\omega^2\tau_{1}^2}{1+\omega^2\tau_{1}^2}= P_{tes,1}^2 |s_I(\omega)|^2 \frac{\omega^2\tau_{1}^2}{1+\omega^2\tau_{1}^2},
\label{TFNnoisehang}
\end{eqnarray}
where we have defined the power TFN spectral densities of the two thermal links as $P_{tes,b}^2$ and $P_{tes,1}^2$, and the total TFN current noise spectral density is simply the sum of the above two components $S_{I,TFN}(\omega)=|I_{\omega}|^2_{tes,b}+|I_{\omega}|^2_{tes,1}$. The exact form of TFN power spectral densities $P_{i}^2$ depend on the details of the nature of the thermal links, and results have been derived for phonon transport in the fully ballistic or 1-D diffusive limits, as  reviewed in Ref. \cite{mccammonbook}, or for electron-phonon or Kapitza thermal conductance \cite{kuzminJAP,dragos}. For all cases where the temperature difference between two adjoining heat capacity blocks is a step function (ballistic phonons, e-p interaction, Kapitza resistance), the TFN power spectral density is well approximated by
\begin{equation}
P_{i,j}^2=2k_B (g_{i,j}(T_{i})T_{i}^2+g_{i,j}(T_{j})T_{j}^2).
\label{powernoise} 
\end{equation}  

Looking at Eqs. \ref{TFNnoisehang}, we see that the first term is the usual TFN noise term ("phonon noise") that exists in the one-block circuit, as well. Its frequency dependence is now simply complicated by the more complex responsivity, and the $|I_{\omega}|^2_{tes,b}$ noise spectra would look just like the responsivity plots in Fig. \ref{SplotH}. On the other hand, the second, new noise term looks very different: It has significant weight only above the frequencies set by the time constant of the extra heat capacity $\tau_{1}$.

\paragraph{Johnson noise of the TES}
\label{johnsonsect}

The Johnson noise of the TES has a slightly more complex influence on the thermal and electrical equations: the noise is electrical in origin, so there is a fluctuation term in the electrical part of the constituent equations. In addition, the dissipated bias power also fluctuates, affecting the thermal part of the circuit. One way to properly take both effects into account was discussed in Ref. \cite{irwinbook}, by introducing the so called internal impedance matrix. We follow the same approach, which leads to a different Taylor expansion of the Joule power (only in terms of current), and to a expression in frequency domain $P_{J}=I_0[(R_0-R_L)-i\omega L]I_{\omega}$, if the simplest bias circuit is assumed. That leads to a set of equations in frequency space
\begin{eqnarray}
i\omega T_{\omega,tes} &=& \frac{(R_0-R_L-i\omega L)I_0}{C_{tes}}I_{\omega}- \frac{1}{\tau_{I}(1-{\cal L}_H)}T_{\omega,tes}+\frac{1}{\tau_{tes,1}}T_{\omega,1} \nonumber\\
i\omega T_{\omega,1} &=& \frac{1}{\tau_{1}}\left ( T_{\omega,tes}-T_{\omega,1} \right ) \nonumber\\
i\omega L I_{\omega} &=& V_{\omega,tes}-\left [ R_{L}+R_{0}(1+\beta) \right ] I_{\omega}-\frac{{\cal L}_{H}(g_{tes,1}+g_{tes,b})}{I_{0}} T_{\omega,tes},
\label{fouriereqs2}
\end{eqnarray}
with the notation the same as for Eqs. \ref{fouriereqs}, and where $V_{\omega,tes}$ is the Johnson voltage noise of the TES, which in equilibrium has the well known expression $V_{\omega,tes}=\sqrt{4k_BT_0R_0}$, but which has been shown \cite{irwinnoise} to depend on the parameter $\beta$ in the first order as $V_{\omega,tes}=\sqrt{4k_BT_0R_0(1+2\beta)}$.

After algebra, one gets an expression for the Johnson current noise spectral density $S_{I,J}(\omega)=|I_{\omega}|^2_{J}$, which is expressed in terms of the known TES complex impedance as
\begin{equation}
|I_{\omega}|^2_{J}=\frac{V_{\omega,tes}^2}{R_{0}^2(2+\beta)^2} \left |\frac{Z_{tes}+R_0}{Z_{circ}} \right |^2,
\label{johnson}
\end{equation}
where we have again used the definition of the circuit impedance $Z_{circ}=Z_{tes}+R_{L}+i\omega L$. This noise adds in quadrature to the TFN noise terms, as it is uncorrelated with them.

\paragraph{Johnson noise of the Thevenin (shunt) resistor, or other external voltage noise}

The Johnson noise due to the shunt and parasitic resistances ($R_L$) does not influence the thermal circuit, and is simply
\begin{equation}
|I_{\omega}|^2_{sh}=\frac{V_{\omega,sh}^2}{|Z_{circ}|^2},
\label{shuntnoise}
\end{equation}
where $Z_{circ}=Z_{tes}+R_{L}+i\omega L$ is the full circuit impedance and 
\begin{equation}
V_{\omega,sh}^2=4k_BT_{sh}R_L, 
\end{equation}
if all of $R_L$ is at temperature $T_{sh}$. If $R_{L}$ consists of two parts, say shunt resistance $R_{sh}$ and parasitic resistance \cite{irwinbook} $R_{para}$, and those two parts are at different temperatures $T_{sh}$ and $T_{para}$, then more accurately  $V_{\omega,sh}^2=4k_B(T_{sh}R_{sh}+T_{para}R_{para})$. Any other external voltage noise would also contribute through Eq. \ref{shuntnoise}, but with $V_{\omega,sh}$ given by the external voltage noise amplitude.

\paragraph{Noise plots}

\begin{figure}[ht]
\includegraphics[width=\textwidth]{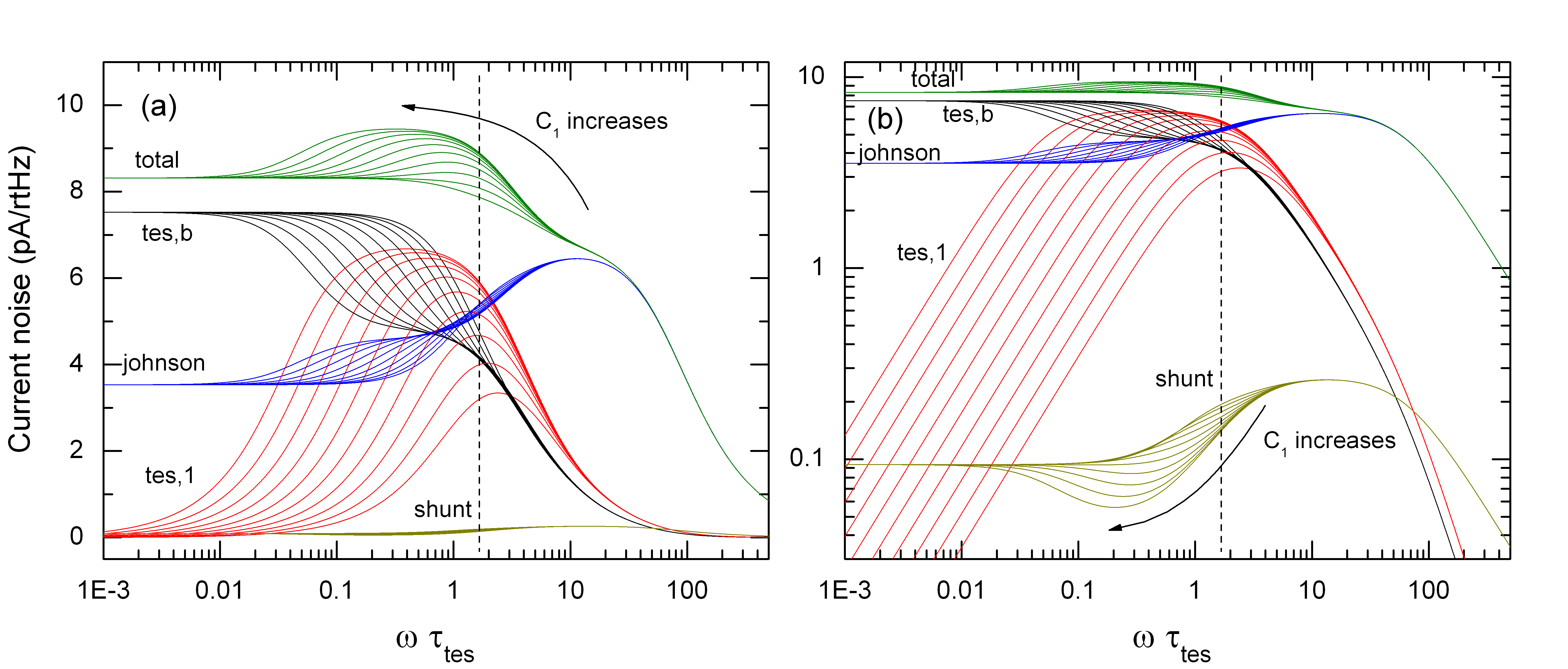}
\caption{(Color online) Current noise of a two-block hanging model, with varying $C_{1}/C_{tes}= 0.33, 0.49, 0.73, 1.1, 1.65, 2.48, 3.71, 5.57, 8.35, 12.5$  in linear-log (a) and log-log (b) scales.   Increasing $C_{1}$ corresponds to a shift of the onset of the bump (or dip in $|I_{\omega}|_{tes,b}$ and $|I_{\omega}|_{sh}$) to lower frequencies, and an increase of the magnitude of $|I_{\omega}|_{tes,1}$ noise component.  Other parameters: ${\cal L}=1.65$, $a=g_{tes,1}/(g_{tes,1}+g_{tes,b})=0.5$,  $R_{0}=0.1 \mathrm{\Omega}$, $\beta=1$,  $R_L=0.001 \mathrm{\Omega}$, $I_0=10 \mu$A, $g_{tes,b}=1$ nW/K, $\tau_{el}=L/[R_L+R_0(1+\beta)]=0.015\tau_{tes}$, $\tau_{tes}=C_{tes}/g_{tes,b}$. Dashed line shows the simple model effective thermal time constant.}
\label{noisetaueffH}       % Give a unique label
\end{figure}

To gain some intuition on the noise, we plot here some examples how the different TES current noise components vary with the thermal parameters. In Fig. \ref{noisetaueffH}, we first show how noise depends on $C_{1}$. When $C_{1}/C_{tes}$ is greater than $\approx 0.5$, a clear bump in the noise spectrum develops in the mid-frequency range between $1/\tau_{1}$ and the effective thermal cut-off set by $C_{tes}$. The onset of the bump moves to lower frequencies with increasing $C_{1}$, and it is, for the used ratio of thermal conductances $a=g_{tes,1}/(g_{tes,1}+g_{tes,b})=0.5$, dominated by the thermal fluctuation noise component of the hanging block $|I_{\omega}|_{tes,1}$, as the phonon noise $|I_{\omega}|_{tes,b}$ actually decreases in the mid-frequency range. The magnitude of the bump saturates to a value determined by the thermal conductance $g_{tes,1}$ when $C_{1}$ is large enough. Johnson noise also develops a minor step feature, and although small in Fig. \ref{noisetaueffH}, its relative strength compared to the TFN noise depends on the details, such as the value of loop gain and resistance.  The shunt resistor noise actually develops a dip for the values of ${\cal L}$ and $a$ used in Fig. \ref{noisetaueffH}, but for the parameter values used here, its magnitude is small.  

\begin{figure}[ht]
\includegraphics[width=\textwidth]{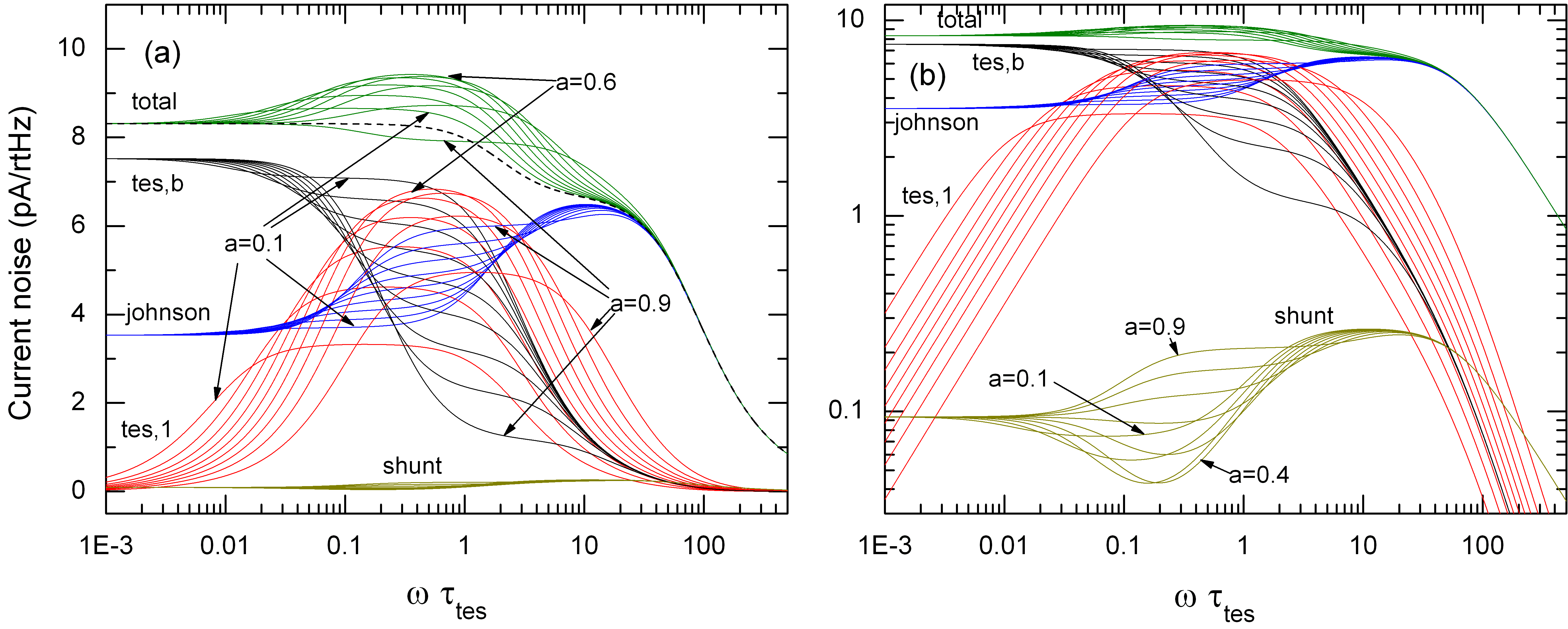}
\caption{ Current noise of a two-block hanging model, with varying $a=g_{tes,1}/(g_{tes,1}+g_{tes,b})= 0.1, 0.2, 0.3, 0.4, 0.5, 0.6, 0.7, 0.8, 0.9$ in linear-log (a) and log-log (b) scales.   Increasing $a$ corresponds to a shift of all the extra features to higher frequencies (both low- and high-frequency sides), a decrease of the phonon noise $|I_{\omega}|_{tes,b}$ and an increase of the Johnson noise $|I_{\omega}|_{J}$ in the mid-frequency range, and a non-monotonous behavior of $|I_{\omega}|_{tes,1}$ and $|I_{\omega}|_{sh}$.  The magnitude of $|I_{\omega}|_{tes,1}$ noise component and the total noise have a maximum at around $a=0.6$, and $|I_{\omega}|_{sh}$ a minimum around $a=0.4$. Other parameters: ${\cal L}=1.65$, $C_{1}/C_{tes}=10$,  $R_{0}=0.1 \mathrm{\Omega}$, $\beta=1$,  $R_L=0.001 \mathrm{\Omega}$, $I_0=10 \mu$A, $g_{tes,b}=1$ nW/K, $\tau_{el}=L/[R_L+R_0(1+\beta)]=0.015\tau_{tes}$, $\tau_{tes}=C_{tes}/g_{tes,b}$. Dashed line in (a) shows the single-block (no $C_{1}$) result for total noise. }
\label{noiseAH}       % Give a unique label
\end{figure}

In Fig. \ref{noiseAH}, the effect of the value of the hanging thermal conductance $g_{tes,1}$ is studied, by plotting the noise spectra as a function of the varying relative strength $a=g_{tes,1}/(g_{tes,1}+g_{tes,b})$. The difference to the effect of $C_{1}$ is that instead of developing a bump with constant magnitude, the bump magnitude evolves with $a$ in a non-monotonous manner, having a maximum at around $a \approx 0.6$. In addition, the bump shifts up in frequency with increasing $a$ (both low and high frequency sides). In other words, the effect of the hanging block on noise is largest for approximately equal magnitudes of the two thermal conductances, just like for the complex impedance. Interestingly, the bump in total noise eventually develops into a step-down feature for the highest values of $a$ here ($a=0.9$), where in a certain region of frequencies the noise can be lower than the simple model noise. This does not imply, however, that the noise equivalent power $NEP=|I_{\omega}|_{tot}/|s_I(\omega)|$ is lower, because the responsivity is reduced even more than noise in that frequency region (See Fig. \ref{SplotH}). Looking at the different components of the noise, one sees that the non-monotonous behavior in total noise is caused by the hanging block TFN noise $|I_{\omega}|_{tes,1}$, whereas the phonon noise $|I_{\omega}|_{tes,b}$ (Johnson noise) decreases (increases) monotonously in mid-frequency range with $a$. The dip in shunt noise in the mid-frequency range is deepest at $a \approx 0.4$, and becomes a step-up feature for $a > 0.6$.     
  
\begin{figure}[ht]
 \includegraphics[width=\textwidth]{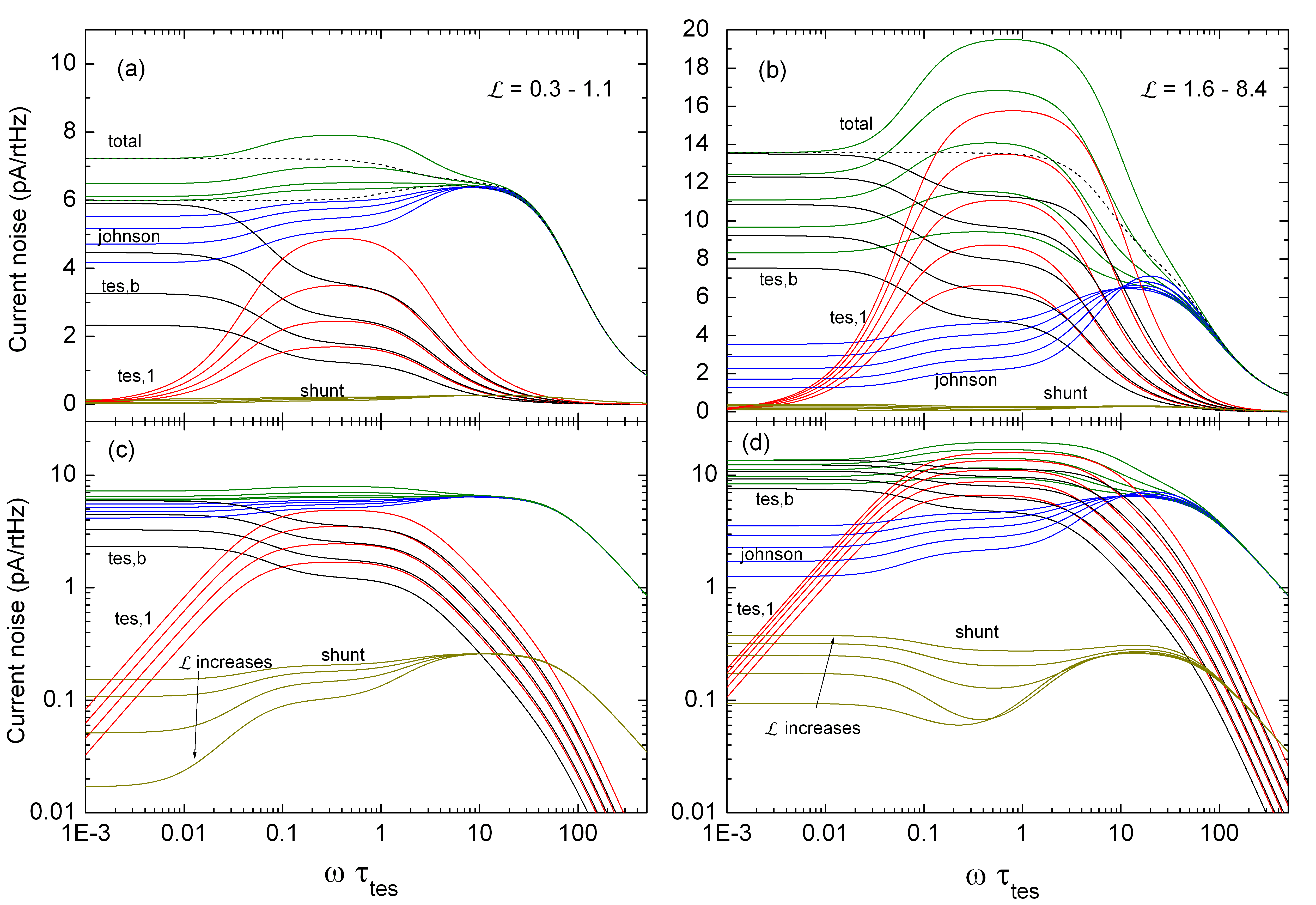}
\caption{(Color online) Current noise of a two-block hanging model, with varying ${\cal L}$. In the left panels ${\cal L} = 0.33, 0.49, 0.73, 1.1$ in linear-log (a) and log-log (c) scales, whereas in the right panels ${\cal L} = 1.65, 2.48, 3.71, 5.57, 8.35 $, in linear-log (b) and log-log (d) scales.   Increasing ${\cal L}$ corresponds to an increase of the $|I_{\omega}|_{tes,b}$ and $|I_{\omega}|_{tes,1}$ TFN noise levels, and an increase of the thermal cut-off frequency for ${\cal L} > 1$. The low-frequency Johnson noise $|I_{\omega}|_{J}$ decreases, and $|I_{\omega}|_{sh}$ behaves non-monotonously.   The minimum  of the low-frequency $|I_{\omega}|_{sh}$ noise component appears at ${\cal L} \approx 1$.  Other parameters: $a=0.5$, $C_{1}/C_{tes}=10$,  $R_{0}=0.1 \mathrm{\Omega}$, $\beta=1$,  $R_L=0.001 \mathrm{\Omega}$, $I_0=10 \mu$A, $g_{tes,b}=1$ nW/K, $\tau_{el}=L/[R_L+R_0(1+\beta)]=0.015\tau_{tes}$, $\tau_{tes}=C_{tes}/g_{tes,b}$. Dashed lines in (a) and (b) show a few results for the the one-block model for total noise. }
\label{noiseLH}       % Give a unique label
\end{figure}

Finally, we also want to plot the dependence on the loop gain ${\cal L}$, shown in Fig. \ref{noiseLH}. The main effect of increasing ${\cal L}$ is the same as in the simple model \cite{irwinbook}, which is that it increases the magnitude of the thermal noise components, and reduces the Johnson noise level below the frequencies set by the effective thermal time constant. This means that for low values of ${\cal L} < 1$ the total noise is Johnson noise limited [Figs \ref{noiseLH} (a) and (c)], whereas for higher loop gains ${\cal L} > 1$ the thermal noise dominates [Figs \ref{noiseLH} (b) and (d)]. This transition means that the relative size of the mid-frequency bump first grows faster with ${\cal L}$, and then more slowly when Johnson noise becomes irrelevant. Notice that for high ${\cal L}$, also the high frequency cut-off for the thermal noise components moves to higher frequencies, due to the increase of the effective thermal time constant. The shunt noise behaves, again, in a more complex way. For low ${\cal L} < 1$, the low-frequency part of it is first suppressed like Johnson noise, but for high ${\cal L} > 1$ the situation reverses, and eventually for the highest values, the low-frequency shunt noise is above the high-frequency level.   

\subsubsection{Intermediate model}

For the intermediate model, equivalent equations to Eqs. \ref{diffeqs2} can be written and solved in frequency domain, with the following simple results for the two TFN noise components:
\begin{eqnarray}
|I_{\omega}|^2_{1,b}&=& P_{1,b}^2 |s_I(\omega)|^2 \frac{g_{tes,1}^2(T_{1})}{(g_{tes,1}(T_{1})+g_{1,b})^2}\frac{1}{1+\omega^2\tau_{1}^2}, \nonumber\\ 
|I_{\omega}|^2_{tes,1}&=& P_{tes,1}^2 |s_I(\omega)|^2 \frac{g_{1,b}^2/(g_{tes,1}(T_{1})+g_{1,b})^2+\omega^2\tau_{1}^2}{1+\omega^2\tau_{1}^2},
\label{TFNnoiseint}
\end{eqnarray}
where $\tau_{1}=C_{1}/(g_{tes,1}(T_{1})+g_{1,b})$, as before for the intermediate model. The Johnson noise results (both TES ans shunt) have no direct dependence on the thermal model parameters, thus Equations \ref{johnson} and \ref{shuntnoise} remain the same, as long as one uses the correct equation for $Z_{tes}$. As $Z_{tes}$ does not qualitatively differ between the intermediate and hanging models, the Johnson noise terms also behave qualitatively the same way. 

\begin{figure}[htp]
\includegraphics[width=\textwidth]{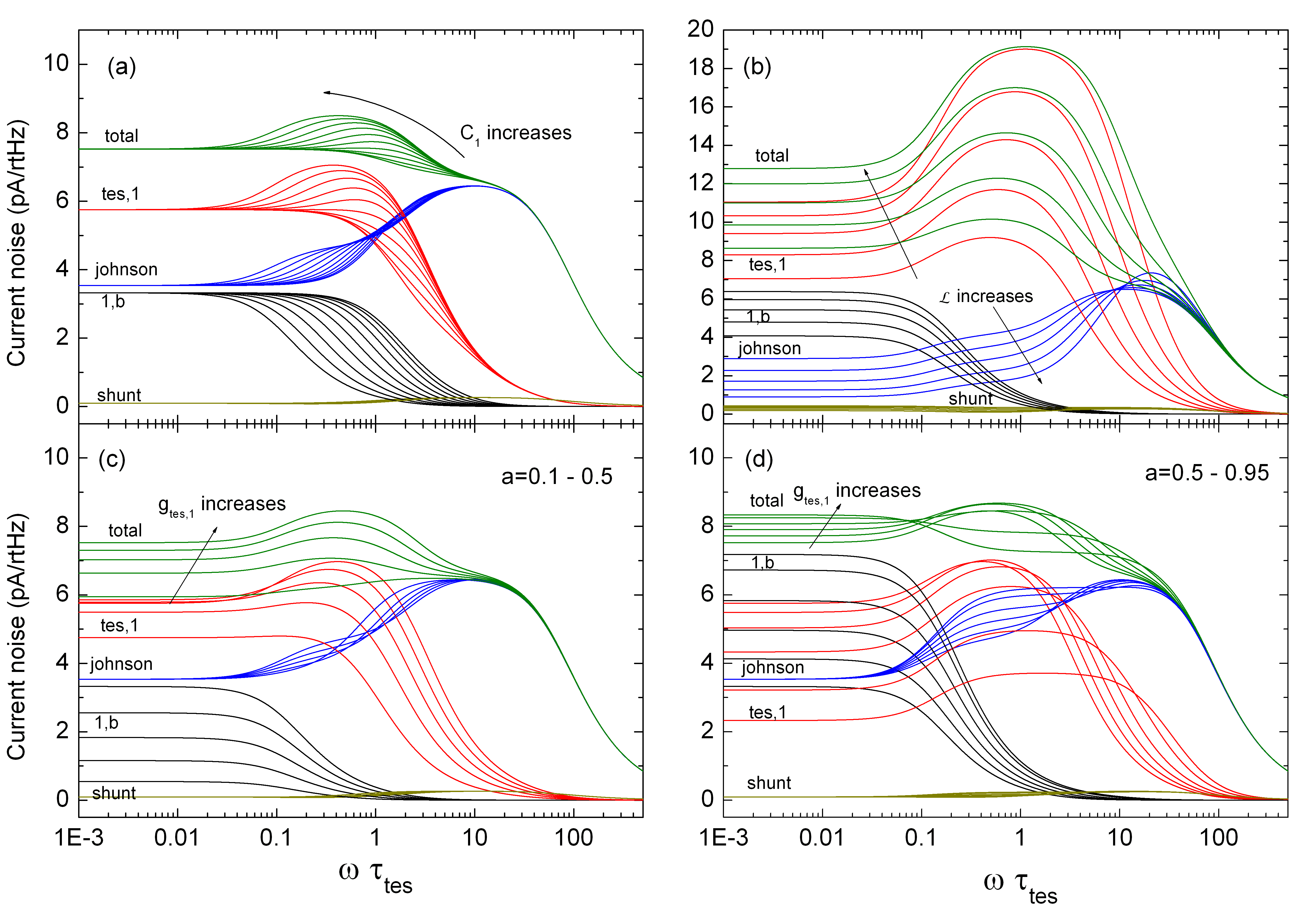}
\caption{ (Color online) (a) Current noise of a two-block intermediate model, with varying $C_{1}/C_{tes}= 0.33, 0.49, 0.73, 1.1, 1.65, 2.48, 3.71, 5.57, 8.35, 12.5$  in linear-log scale. Increasing $C_{1}$ corresponds to a growth of the bump and a shift of its onset to lower frequencies, and a decrease of the cut-off frequency for the $|I_{\omega}|_{1,b}$ noise component. (b) Same, with varying ${\cal L} = 2.48, 3.71, 5.57, 8.35, 12.5$. Increasing ${\cal L}$ corresponds to an increase of the $|I_{\omega}|_{1,b}$ and $|I_{\omega}|_{tes,1}$ noise levels,  and an increase of the thermal cut-off frequency for $|I_{\omega}|_{tes,1}$. (c) Same, with varying   $a=g_{tes,1}(T_{1})/(g_{tes,1}(T_{1})+g_{1,b})= 0.1, 0.2, 0.3, 0.4, 0.5$, and (d) with $a= 0.5, 0.6, 0.7, 0.8, 0.9, 0.95$.  Increasing $a$ corresponds to a monotonous increase of the phonon noise level $|I_{\omega}|_{1,b}$, but a non-monotonous behavior of $|I_{\omega}|_{tes,1}$ [first increase in (c), then decrease in (d)] with maximum bump amplitude around $a=0.6$ (maximum low-frequency noise around $a=0.4$), and a shift of the bump to higher frequencies.  Johnson noise $|I_{\omega}|_{J}$ increases in the mid-frequency range. Other parameters in plots (if not varied): ${\cal L}=1.65$, $C_{1}/C_{tes}=10$,  $a=0.5$, $R_{0}=0.1 \mathrm{\Omega}$, $\beta=1$,  $R_L=0.001 \mathrm{\Omega}$, $I_0=10 \mu$A, $g_{1,b}=1$ nW/K, $\tau_{el}=L/[R_L+R_0(1+\beta)]=0.015\tau_{tes}$, $\tau_{tes}=C_{tes}/g_{1,b}$. }
\label{noiseITF}       % Give a unique label
\end{figure}

In Fig. \ref{noiseITF} we plot examples of how the noise depends on the most relevant parameters for this model. $T_{1}$ was calculated using the simplifying assumption $n=m=4$. The results for the total noise as a function of $C_{1}$ [Fig. \ref{noiseITF} (a)] look nearly identical to the results of the hanging model, Fig. \ref{noisetaueffH}. The only main difference is that the level of low-frequency noise is lower because the intermediate block is at a lower temperature than the hanging block, thus reducing the TFN noise level. Naturally, the breakdown of the TFN noise into two components is completely different in this case: both $|I_{\omega}|_{1,b}$ and $|I_{\omega}|_{tes,1}$ have a low-frequency component, with $|I_{\omega}|_{tes,1}$ developing the bump structure at intermediate frequencies above $1/\tau_{1}$. Fig \ref{noiseITF} (b) shows the dependence on ${\cal L}$ for high ${\cal L}$ values. Again, the total noise is nearly identical. Interestingly, in the intermediate model the shape of $|I_{\omega}|_{tes,1}$ noise actually changes with ${\cal L}$, by increasing the relative size of the bump-feature. The result of that is that the total TFN noise looks nearly the same as in the hanging model, in which the shape changes due to the stronger growth of the hanging noise component relative to the phonon noise.

In Figs.  \ref{noiseITF} (c) and (d) we show the dependence on $g_{tes,1}$, keeping $g_{1,b}$ constant, parametrized by the relative strength of the two thermal conductances, $a=g_{tes,1}(T_{1})/(g_{tes,1}(T_{1})+g_{1,b})$. Once again, the dependence is more complicated. The total noise   has a maximum at $a \approx 0.6-0.7$ at intermediate frequencies (the bump structure), and the bump shifs up in frequency with increasing $a$, and finally vanishes for high enough $a$, as before for the hanging model. The low frequency noise increases because of the increasing effective conductance to the bath, in contrast to the hanging case. By looking at the TFN noise components separately, once sees that the appearance and disappearance of the bump is entirely due to the $|I_{\omega}|_{tes,1}$ component (having a maximum around $a=0.5$), as  the $|I_{\omega}|_{1,b}$ part simply increases monotonously with $a$, becoming dominant at low frequencies for high $a$. The Johnson noise again increases in the mid-frequency range with $a$. For the highest value of $a$ here, the total noise already looks like the noise of a single-block model, but with a larger heat capacity $C_{tes}+C_{1}$.

\subsubsection{Parallel model}

Finally, for the parallel model [Fig. \ref{twoblocks} (c)], due to the one extra thermal link compared to the hanging and intermediate models, there will be one more TFN noise component. Solving for all the TFN noise components, one gets:
\begin{eqnarray}
|I_{\omega}|^2_{1,b}&=& P_{1,b}^2 |s_I(\omega)|^2 \frac{g_{tes,1}^2(T_{1})}{(g_{tes,1}(T_{1})+g_{1,b})^2}\frac{1}{1+\omega^2\tau_{1}^2}, \nonumber\\ 
|I_{\omega}|^2_{tes,1}&=& P_{tes,1}^2 |s_I(\omega)|^2 \frac{g_{1,b}^2/(g_{tes,1}(T_{1})+g_{1,b})^2+\omega^2\tau_{1}^2}{1+\omega^2\tau_{1}^2}, \nonumber\\
|I_{\omega}|^2_{tes,b}&=& P_{tes,b}^2 |s_I(\omega)|^2,
\label{TFNnoiseparallel}
\end{eqnarray}
with $\tau_{1}=C_{1}/(g_{tes,1}(T_{1})+g_{1,b})$ (Eq. \ref{parasymbols}) the same as for the intermediate model, because the added $g_{tes,b}$ does not contact $C_{1}$.

\begin{figure}[ht]
\includegraphics[width=\textwidth]{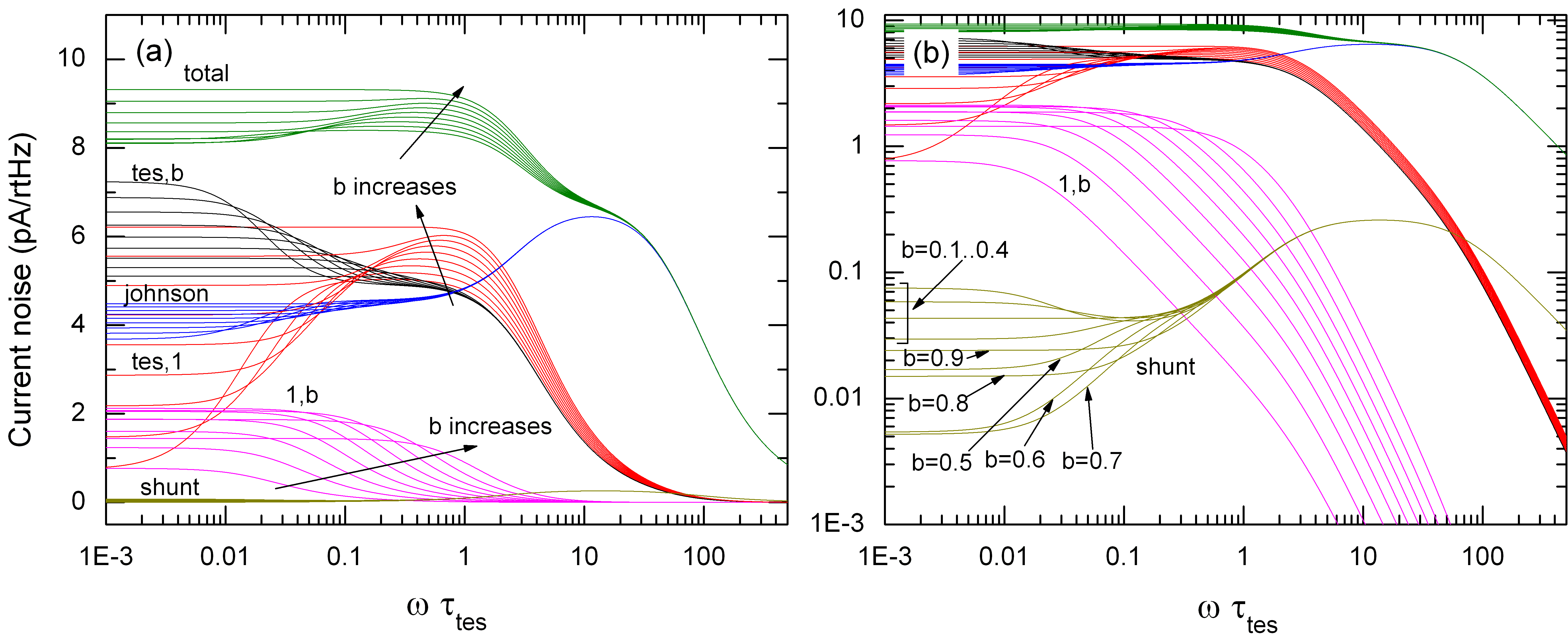}
\caption{(Color online) Current noise of the two-block parallel model, with varying $b=g_{1,b}/[g_{1,b}+g_{tes,1}(T_{1})]=0.1,0.2,0.3,0.4,0.5,0.6,0.7,0.8,0.9$  in (a) linear-log and (b) log-log scales. Increasing $b$ corresponds to a decrease of the low-frequency $|I_{\omega}|_{tes,b}$ noise, increase of the low-frequency $|I_{\omega}|_{tes,1}$ noise and a non-monotonous behavior of the $|I_{\omega}|_{1,b}$ noise, with a maximum around $b=0.5$. Low-frequency Johnson noise increases, and shunt noise has a minimum around $b=0.7$.  Other parameters: ${\cal L}=1.65$, $C_{1}/C_{tes}=10$,  $a=g_{tes,1}/(g_{tes,1}+g_{tes,b})=0.5$, $R_{0}=0.1 \mathrm{\Omega}$, $\beta=1$,  $R_L=0.001 \mathrm{\Omega}$, $I_0=10 \mu$A, $g_{1,b}=1$ nW/K, $\tau_{el}=L/[R_L+R_0(1+\beta)]=0.015\tau_{tes}$, $\tau_{tes}=C_{tes}/g_{tes,b}$.}
\label{noiseparallel}       % Give a unique label
\end{figure} 

In Fig. \ref{noiseparallel}, we show how the different noise components evolve as a function of the thermal conductance $g_{1,b}$. This means that the starting point is the hanging model with $g_{1,b}=0$. Again, $T_{1}$ was calculated using the simplifying assumption $n=m=4$. The general picture is that the total noise increases as $g_{1,b}$ increases, parametrized in the plot as $b=g_{1,b}/[g_{1,b}+g_{tes,1}(T_{1})]$. In terms of the noise components, the low-frequency shoulder of the phonon noise $|I_{\omega}|_{tes,b}$ decreases, and when $b > 0.8$, the phonon noise becomes flat up to the thermal cut-off frequency.  On the other hand, the hanging TFN noise $|I_{\omega}|_{tes,1}$ develops an increasing low-frequency level (as in the intermediate model) quite fast, eventually surpassing the low frequency noise contribution from the phonon noise at around $b = 0.7-0.8$. In addition, the intermediate frequency bump size also increases with $b$, and moves to higher frequencies. The TFN noise component produced by the direct coupling of $C_{1}$ to bath, $|I_{\omega}|_{1,b}$, has only a low frequency component, whose cut-off moves up in frequency monotonously with $b$. Also, the strength of the cut-off for $|I_{\omega}|_{1,b}$ increases with $b$, as can be seen from the log-log plot, Fig. \ref{noiseparallel} (b). This is in contrast to the other two TFN noise components. The noise level for $|I_{\omega}|_{1,b}$  initially increases, but then reaches a maximum around $b=0.5$, after which it starts decreasing. For the parameter values chosen here, this last TFN component $|I_{\omega}|_{1,b}$ is the smallest, but not insignificant. The low-frequency Johnson noise level increases with $b$, and the shunt noise behaves in a complex manner at low frequencies, where it has the highest value at low $b$, decreases until $b=0.7$, and then starts increasing again for $b > 0.7$ [See Fig. \ref{noiseparallel} (b)]. Finally,  in the intermediate bump region the total noise increases monotonously with $b$. However, for low frequencies the total noise initially {\em decreases} slightly and has a minimum at $b=0.2-0.3$ (for the parameter values used), after which the low-frequency part also increases.

\section{Three-block models}
\label{sec:2}

\begin{figure}[ht]
  \includegraphics[width=0.7\textwidth]{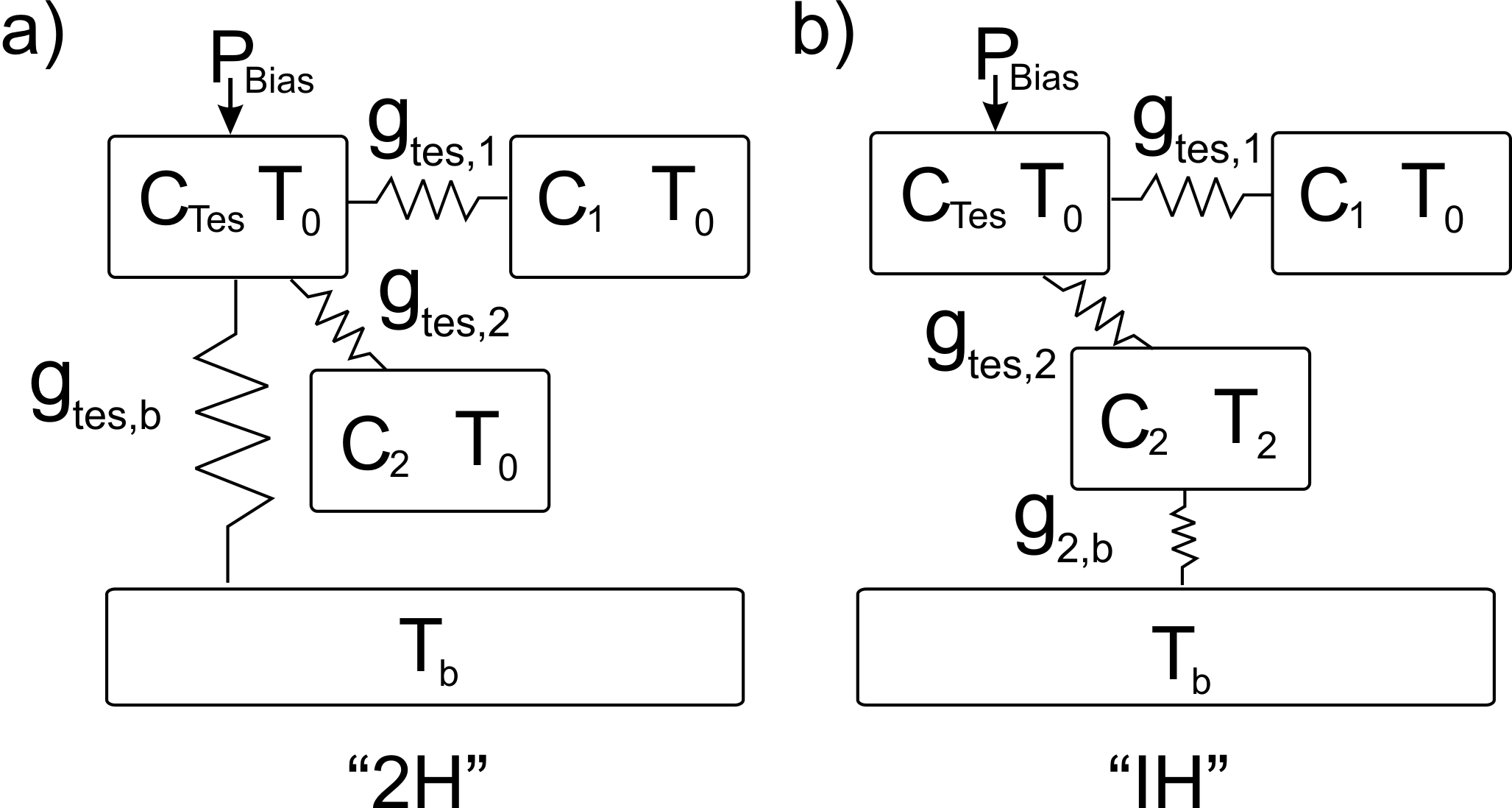}
\caption{Three-block models studied. (a) 2H model, (b) IH model.}
\label{3blockmodels}       % Give a unique label
\end{figure}

After the exhaustive discussion of the two-block models, we limit ourselves here to two examples of three-block models, which we have already used in analysis of real TES data \cite{mikko,kimmo}, see Fig. \ref{3blockmodels}. The first model is the analog of the hanging two-block model, where there are now two hanging extra heat capacities $C_{1}$ and $C_{2}$. We have named this the 2H model. The second model has one intermediate and one hanging block, and we name it the IH model. 

\subsection{Equations to be solved}

The extension of the differential equations defining the problem for two-block models, Eq. \ref{diffeqs}, to the equations for three-block models is straightforward. One more equation is generated because of the new heat capacity, and one more power flow term is added, which in the case of the 2H model is $C(T_{tes}^p-T_{2}^p)$ in the equation describing $C_{tes}$. 
Then, after linearization and transformation to frequency domain, the set of equations become for the 2H model
\begin{eqnarray}
i\omega T_{\omega,tes} &=& \frac{I_{0}R_{0}(2+\beta)}{C_{tes}}I_{\omega}- \frac{1}{\tau_I}T_{\omega,tes}+\frac{1}{\tau_{tes,1}}T_{\omega,1}+\frac{1}{\tau_{tes,2}}T_{\omega,2}+\frac{1}{C_{tes}}P_{\omega}, \nonumber\\
i\omega T_{\omega,1} &=& \frac{1}{\tau_{1}}\left ( T_{\omega,tes}-T_{\omega,1} \right ), \nonumber\\
i\omega T_{\omega,2} &=& \frac{1}{\tau_{2}}\left ( T_{\omega,tes}-T_{\omega,2} \right ), \nonumber\\
i\omega L I_{\omega} &=& V_{\omega}-\left [ R_{L}+R_{0}(1+\beta) \right ] I_{\omega}-\frac{{\cal L}_{2H}(g_{tes,1}+g_{tes,2}+g_{tes,b})}{I_{0}} T_{\omega,tes},
\label{fouriereqs2H}
\end{eqnarray}
where all $g_i$ are evaluated at $T_0$, $\tau_{tes,1} = C_{tes}/g_{tes,1}$, $\tau_{tes,2} = C_{tes}/g_{tes,2}$, and the definitions of the more important time constants $\tau_i$ and the effective loop gain ${\cal L}_{2H}$ are now:
\begin{eqnarray}
\tau_{I} &=& \frac{C_{tes}}{(g_{tes,1}+g_{tes,2}+g_{tes,b})(1-{\cal L}_{2H})}, \nonumber\\
\tau_{1} &=& \frac{C_{1}}{g_{tes,1}}, \nonumber\\
\tau_{2} &=& \frac{C_{2}}{g_{tes,2}}, \nonumber\\
{\cal L}_{2H} &=& \frac{P_{0}\alpha}{(g_{tes,1}+g_{tes,2}+g_{tes,b})T_{0}}.
\label{2Hdefs}
\end{eqnarray}

Comparing to the two-block equations \ref{fouriereqs}, we see that the first equation has one more term due to the added heat capacity $C_{2}$, the third equation is new, and the last term in the equation for $I_{\omega}$ has a changed coefficient due to the new thermal conductance $g_{tes,2}$. 
  
\subsection{Complex impedance}

\subsubsection{2H model}

\begin{figure}[p]
\includegraphics[width=0.9\textwidth]{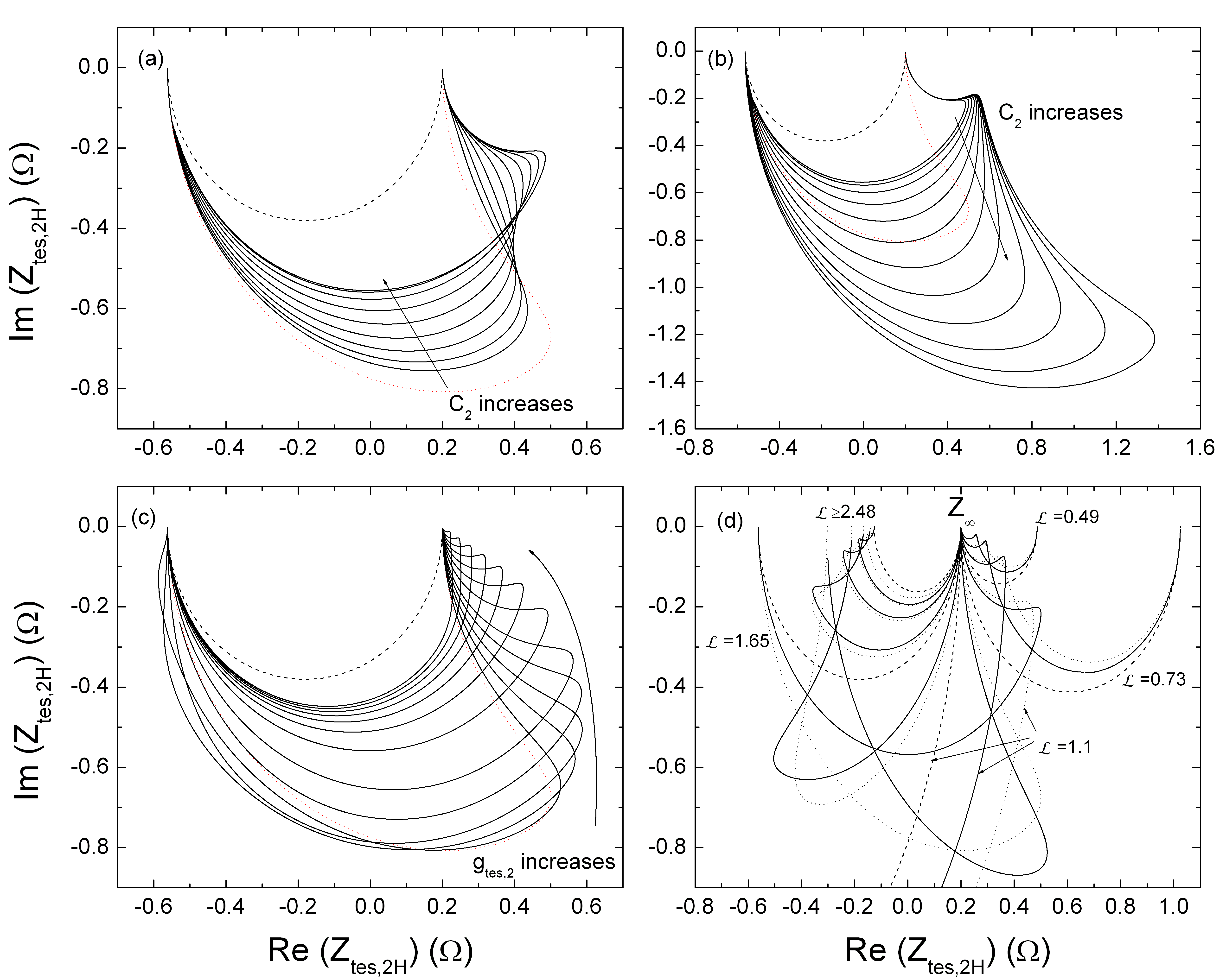}
\caption{ Complex impedance of the three-block 2H model, with varying parameters. We have used $R_{0}=0.1 \mathrm{\Omega}$ and $\beta=1$ in all plots, and frequencies run typically between $\omega\tau_{tes} = 0.01 .. 100$ (for some curves $10^{-4}..100$), with $\tau_{tes}=C_{tes}/g_{tes,b}$.  (a) $Z_{tes,2H}$ as a function of $C_{2}$, with $C_{2}/C_{tes}= 0.33, 0.49, 0.73, 1.1, 1.65, 2.48, 3.71, 5.57, 8.35$, $C_{1}/C_{tes}=10$, ${\cal L}=1.65$, $a=g_{tes,1}/(g_{tes,1}+g_{tes,b})=0.5$ and $d_{2}=g_{tes,2}/(g_{tes,1}+g_{tes,2}+g_{tes,b})=0.3$. Increasing $C_{2}$ corresponds to a shrinking of the bulge caused by $C_{1}$, and a growth of a second one at higher frequencies. The two-block model limit ($C_{2}=0$) is shown as the red dotted line, and the one-block limit ($C_{1}=0$) as the dashed line. (b) Same, but with $C_{2}/C_{tes}= 8.35, 12.5, 18.8, 28.2, 42.3, 63.4, 95.1, 142.7, 214.1, 321.1, 481.7, 722.5$. Increasing $C_{2}$ corresponds to a growth of the bulge. The dotted and dashed lines are the same as in (a). (c)  $Z_{tes,2H}$ as a function of $d_{2}$ (or $g_{tes,2}$), with $d_{2}=0.02, 0.06, 0.1, 0.15, 0.2, 0.3, 0.4, 0.5, 0.6, 0.7, 0.8, 0.9$, ${\cal L}=1.65$, $C_{1}/C_{tes}=10$, $C_{2}/C_{tes}=10$, and $a=0.5$. Increasing $d_{2}$ corresponds to the added bulge feature moving from the left, low-frequency side to the right, high-frequency side. The dotted and dashed lines are the same as in (a). (d) $Z_{tes,2H}$ as a function of ${\cal L} = P_{0}\alpha/(g_{tes,b}T_{0})$, with ${\cal L}= 0.49, 0.73, 1.1, 1.65, 2.48, 3.71, 5.57, 8.35, 12.5$, $C_{1}/C_{tes}=10$, $C_{2}/C_{tes}=10$, $d_{2}=0.3$ and $a=0.5$. The two-block limits ($C_{2}=0$) are shown as dotted lines, and some one-block results as dashed lines. }
\label{plots2H}       % Give a unique label
\end{figure}

The complex impedance for the 2H model is calculated from Eqs. \ref{fouriereqs2H} the same way as for the two-block models, with the result
\begin{equation}
%\begin{eqnarray}
Z_{tes,2H}=R_0(1+\beta)+\frac{{\cal L}_{2H}}{1-{\cal L}_{2H}}R_0(2+\beta) \left /  \left [1+i\omega\tau_{I}-\frac{d_{1}}{1-{\cal L}_{2H}}\frac{1}{1+i\omega\tau_{1}}
-\frac{d_{2}}{1-{\cal L}_{2H}}\frac{1}{1+i\omega\tau_{2}} \right ] \right.,
\label{Z2H}
\end{equation}
%\end{eqnarray}
where we have denoted the relative strengths of the thermal conductances as $d_{1}=g_{tes,1}/(g_{tes,1}+g_{tes,2}+g_{tes,b})$ and $d_{2}=g_{tes,2}/(g_{tes,1}+g_{tes,2}+g_{tes,b})$, and other symbols are defined in Eqs. \ref{2Hdefs}.
By comparing with the result for the hanging two-block model, Eq. \ref{Zhang}, we see that because of the added $C_{2}$, a new term appears in the denominator. This term is naturally mathematically equivalent with the term for $C_{1}$, as both $C_{1}$ and $C_{2}$ are hanging in this model. However, one should bear in mind that the term for $C_{1}$ is not exactly the same as in the hanging two-block model, because the new thermal conductance $g_{tes,2}$ also affects it through the pre-factor $d_{1}$. Also, one sees from Eq. \ref{Z2H} that the effect of $C_{2}$ is not additive, as the new term is in the denominator. It is therefore not obvious how $Z_{tes,2H}$ behaves as a function of the new thermal parameters $C_{2}$ and $g_{tes,2}$, and we therefore investigate their effect by examples, shown in Fig. \ref{plots2H}.   

What we observe from Figs. \ref{plots2H} (a) and (b) is that by increasing $C_{2}$, the bulge feature caused by $C_{1}$ first gets smaller, and starts to be pushed to a higher frequency (further right). However, when $C_{2} \approx C_{1}$, the trend turns and the bulge feature at the original place also starts increasing again. Importantly, for $C_{2} \le C_{1}$ the new term does not necessarily lead to a new, additive bulge in the complex plane, but just contributes to the old, existing one by changing its shape. As a function of $d_{2}=g_{tes,2}/(g_{tes,1}+g_{tes,2}+g_{tes,b})$ [Fig. \ref{plots2H} (c)] , things look a bit like for parameter $a$ in the two-block case. Interestingly, there is always an increased bulge compared to the two-block model on the high frequency side, but also on low frequency side for small $d_{2} < 0.15$ for these parameter values.   
In the limit $d_{2} \rightarrow 1$ $Z_{tes,2H}$ approaches a two-block model with a higher effective TES heat capacity $C_{tes}+C_{2}$. The loop gain again has a complex effect, as shown in Fig. \ref{plotsH} (d). In the large loop gain limit, the three-block hanging model approaches the simple model, but more slowly than in the two-block case. For typical values of  ${\cal L}$ the effect of the added term is strong, as can be seen by comparing the two- and three block results.  

\subsubsection{IH model}

The starting equations for the IH model look just slightly different from the 2H case, as the thermal conductances couple different heat capacities, which can be at different steady state temperatures ($T_{0}$ for $C_{tes}$ and $C_{1}$, and $T_{2}$ for $C_{2}$). The derivation thus follows analogously to the 2H model, and we simply quote the end result for the complex impedance.
It looks mathematically exactly the same as the result for the 2H model, with only some changes in parametrization:
\begin{equation}
Z_{tes,IH}=R_0(1+\beta)+\frac{{\cal L}_{IH}}{1-{\cal L}_{IH}}R_0(2+\beta) \left / \left [1+i\omega\tau_{I}-\frac{d_{1}}{1-{\cal L}_{IH}}\frac{1}{1+i\omega\tau_{1}}
-\frac{d_{2}}{1-{\cal L}_{IH}}\frac{1}{1+i\omega\tau_{2}} \right ] \right.,
\label{ZIH}
\end{equation}
where the definitions of the prefactors $d_i$, the time constants $\tau_i$ and the effective loop gain ${\cal L}_{IH}$ are now:
\begin{eqnarray}
d_{1}&=&\frac{g_{tes,1}}{g_{tes,1}+g_{tes,2}(T_{0})}  \nonumber\\
d_{2}&=&\frac{g_{tes,2}(T_{0})g_{tes,2}(T_{2})}{(g_{tes,1}+g_{tes,2}(T_{0}))(g_{tes,2}(T_{2})+g_{2,b})} \nonumber\\
\tau_{I} &=& \frac{C_{tes}}{[g_{tes,1}+g_{tes,2}(T_{0})](1-{\cal L}_{IH})}, \nonumber\\
\tau_{1} &=& \frac{C_{1}}{g_{tes,1}}, \nonumber\\
\tau_{2} &=& \frac{C_{2}}{g_{tes,2}(T_{2})+g_{2,b}}, \nonumber\\
{\cal L}_{IH} &=& \frac{P_{0}\alpha}{[g_{tes,1}+g_{tes,2}(T_{0})]T_{0}}.
\end{eqnarray}

%\tau_{tes,1} &=& \frac{C_{tes}}{g_{tes,1}}, \nonumber\\
%\tau_{tes,2} &=& \frac{C_{tes}}{g_{tes,2}(T_{2})}, \nonumber\\
%\tau_{2b} &=& \frac{C_{2}}{g_{tes,2}(T_{0})}, \nonumber\\
%\begin{eqnarray}
%i\omega T_{\omega,tes} &=& \frac{I_{0}R_{0}(2+\beta)}{C_{tes}}I_{\omega}- \frac{1}{\tau_I}T_{\omega,tes}+\frac{1}{\tau_{tes,1}}T_{\omega,1}+\frac{1}{\tau_{tes,2}}T_{\omega,2}+P_{\omega}, \nonumber\\
%i\omega T_{\omega,1} &=& \frac{1}{\tau_{1}}\left ( T_{\omega,tes}-T_{\omega,1} \right ), \nonumber\\
%i\omega T_{\omega,2} &=& \frac{1}{\tau_{2b}} T_{\omega,tes}-\frac{1}{\tau_{2}} T_{\omega,2}, \nonumber\\
%i\omega L I_{\omega} &=& V_{\omega}-\left [ R_{L}+R_{0}(1+\beta) \right ] I_{\omega}-\frac{{\cal L}_{IH}}{(g_{tes,1}+g_{tes,2}(T_{0}))}{I_{0}} T_{\omega,tes},
%\label{fouriereqsIH}
%\end{eqnarray}
$g_{tes,1}$ is always evaluated at $T_{0}$ and $g_{2,b}$ at $T_{2}$, therefore we have omitted the temperature dependence from their notation for simplification. $g_{2,b}(T_{b})$ will only come into play through Eq. \ref{powernoise}, when calculating the noise amplitude. The true loop gain is again the same as for the intermediate two-block model (if one takes into account the notation change $C_{1} \rightarrow C_{2}$), given by ${\cal L}=P_{0}\alpha/(g_{eff}T_{0})$, where 
\begin{equation}
g_{eff}=\frac{g_{tes,2}(T_{0})g_{2,b}}{g_{tes,2}(T_{2})+g_{2,b}}. 
\end{equation} 
Because of the mathematical equivalence between $Z_{tes,2H}$ and $Z_{tes,IH}$, we do not discuss the details further here, similar plots to Fig. \ref{plots2H} could be generated.

\subsection{Small-signal responsivity}

\begin{figure}[ht]
\includegraphics[width=\textwidth]{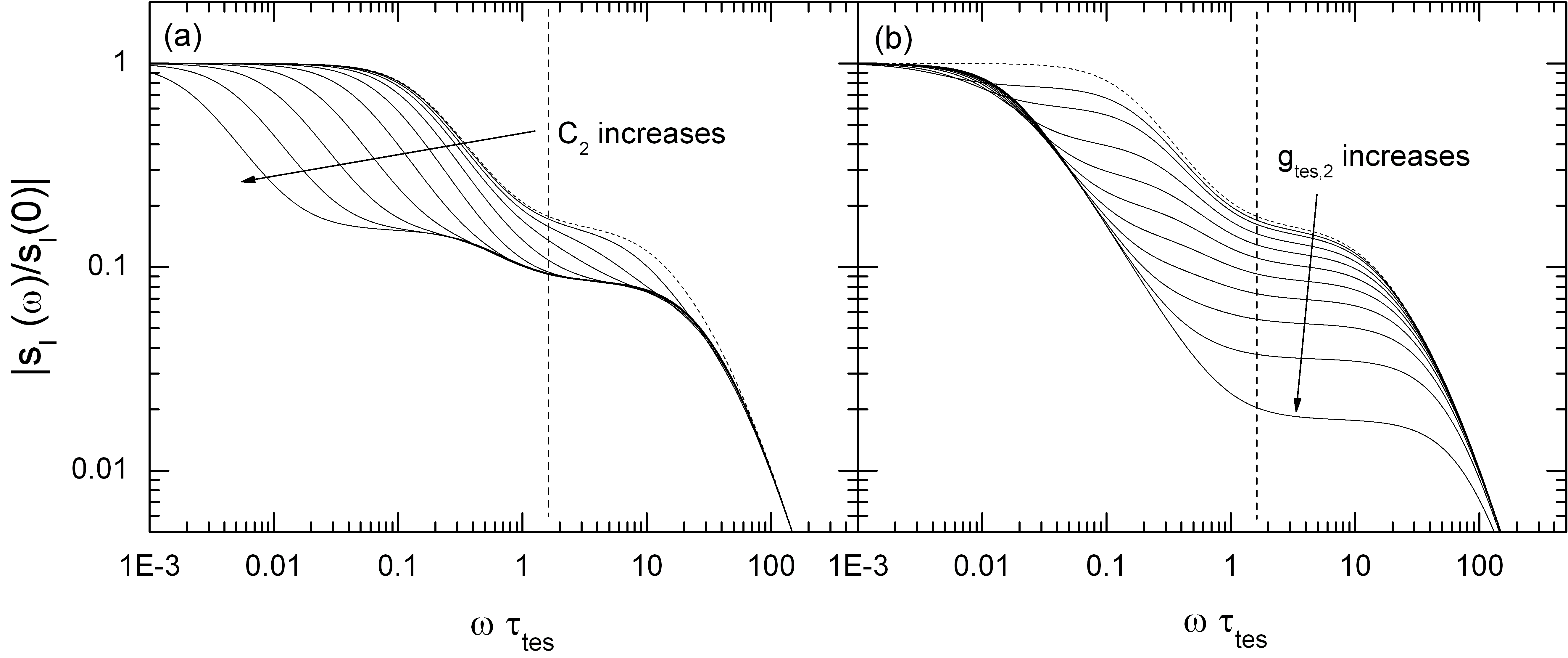}
\caption{Responsivity of a three-block 2H model, with varying thermal parameters. (a) $|s_I(\omega)|$ as a function of $C_{2}$, with $C_{2}/C_{tes}= 0.33, 1.1, 2.48, 5.57, 12.5, 28.2, 63.4, 142.7, 321.1, 722.5$, and $d_{2}=g_{tes,2}/(g_{tes,1}+g_{tes,2}+g_{tes,b})=0.5$. Increasing $C_{2}$ corresponds to a shift of the first partial thermal cut-off to lower frequencies and the development of a "double knee" in the intermediate frequency range. (b) $|s_I(\omega)|$  as a function of $d_{2}$ (or $g_{tes,2}$), with $d_{2}=0.05,0.1, 0.2, 0.3, 0.4, 0.5, 0.6, 0.7, 0.8, 0.9$, $C_{2}/C_{tes}=100$. Increasing $d_{2}$ corresponds to the decrease of responsivity. We have used $R_{0}=0.1 \mathrm{\Omega}$, $\beta=1$, $\tau_{el}=L/[R_L+R_0(1+\beta)]=0.015\tau_{tes}$, $C_{1}/C_{tes}=10$, $a=g_{tes,1}/(g_{tes,1}+g_{tes,b})=0.9$ and ${\cal L}=1.65$ in all plots (if not varied). $\tau_{tes}=C_{tes}/g_{tes,b}$. Dashed vertical line shows the simple model effective thermal time constant, and dotted curves the corresponding two-block results (limit $C_{2}=0$). }
\label{Splot2H}       % Give a unique label
\end{figure}

Again, derivation of the linear responsivity is straightforward using Eqs. \ref{fouriereqs2H} or the equivalent ones for the IH model. As the model dependent terms can all be lumped into $Z_{tes}$, both three-block models also satisfy Equation \ref{respo}. In Fig. \ref{Splot2H} we plot examples of how $|s_I(\omega)|$ behaves for the 2H model, similar plots would also follow for the IH model. The overall behavior is the same as with the two-block models, with a partial thermal cut-off moving to lower frequencies with increasing $C_{2}$, and a decrease of responsivity above that partial thermal cut-off with increasing $d_{2}=g_{tes,2}/(g_{tes,1}+g_{tes,2}+g_{tes,b})$. A notable difference to the two-block results is that for some values of the parameters $C_{2}$ and $d_{2}$ [$C_{2} > C_{1}$ in Fig. \ref{Splot2H} (a) and $d_{2} < 0.6 $ in Fig. \ref{Splot2H} (b)], the responsivity shows a "double knee" structure.   

\subsection{Thermodynamic noise}

\subsubsection{2H model}

\paragraph{Thermal fluctuation noise}

The thermal fluctuation noise components for the 2H model can be derived analogously to the two-block models, by adding the power fluctuation terms to equations  \ref{fouriereqs2H}. Straightforward algebra then yields for the three TFN noise components
\begin{eqnarray}
|I_{\omega}|^2_{tes,1}&=& P_{tes,1}^2 |s_I(\omega)|^2 \frac{\omega^2\tau_{1}^2}{1+\omega^2\tau_{1}^2}, \nonumber\\
|I_{\omega}|^2_{tes,2}&=& P_{tes,2}^2 |s_I(\omega)|^2 \frac{\omega^2\tau_{2}^2}{1+\omega^2\tau_{2}^2}, \nonumber\\
|I_{\omega}|^2_{tes,b}&=& P_{tes,b}^2 |s_I(\omega)|^2,
\label{TFNnoise2H}
\end{eqnarray}
if we assume that the three TFN noise sources are all uncorrelated with each other. The power noise amplitudes $P_i$ are defined as before by Eq. \ref{powernoise}, and the time constants as $\tau_i=C_i/g_{tes,i}$, just as for the impedance in Eqs. \ref{2Hdefs}.

\paragraph{Johnson noise terms}

Again, we can derive the TES Johnson noise using the internal impedance matrix formulation of section \ref{hangsect}. Just as in the case of responsivity, the end result is that the thermal model dependency is fully accounted for by the TES complex impedance, so that Eq. \ref{johnson} is still valid. The same applies to the external (shunt) Johnson noise term, where Eq. \ref{shuntnoise} can still be used.

We skip plotting the noise for the 2H model, as Eqs. \ref{TFNnoise2H} clearly show that the overall picture is analogous to the two-block hanging model. Now, we simply have two independent TFN noise bumps, which are multiplied by the responsivity curves shown in Fig. \ref{Splot2H}. Because of the monotonously decreasing shape of $|s_I(\omega)|$, the higher frequency TFN noise bump is always suppressed more than the lower frequency bump.  
 
\subsubsection{IH model}

For the IH model, the TFN current noise terms are naturally derived similarily, with the result
\begin{eqnarray}
|I_{\omega}|^2_{tes,1}&=& P_{tes,1}^2 |s_I(\omega)|^2 \frac{\omega^2\tau_{1}^2}{1+\omega^2\tau_{1}^2}, \nonumber\\
|I_{\omega}|^2_{tes,2}&=& P_{tes,1}^2 |s_I(\omega)|^2 \frac{g_{2,b}^2/(g_{tes,2}(T_{2})+g_{2,b})^2+\omega^2\tau_{2}^2}{1+\omega^2\tau_{2}^2}, \nonumber\\
|I_{\omega}|^2_{2,b}&=& P_{2,b}^2 |s_I(\omega)|^2 \frac{g_{tes,2}^2(T_{2})}{(g_{tes,2}(T_{2})+g_{2,b})^2}\frac{1}{1+\omega^2\tau_{2}^2},  
\label{TFNnoiseIH}
\end{eqnarray}
where $\tau_{1}=C_{1}/g_{tes,1}$ and $\tau_{2}=C_{2}/(g_{tes,2}(T_{2})+g_{2,b})$, as before for the IH model. The Johnson noise for the TES and the shunt follow again from the general formulas Eq. \ref{johnson} and Eq. \ref{shuntnoise}.

\begin{figure}[p]
\includegraphics[width=\textwidth]{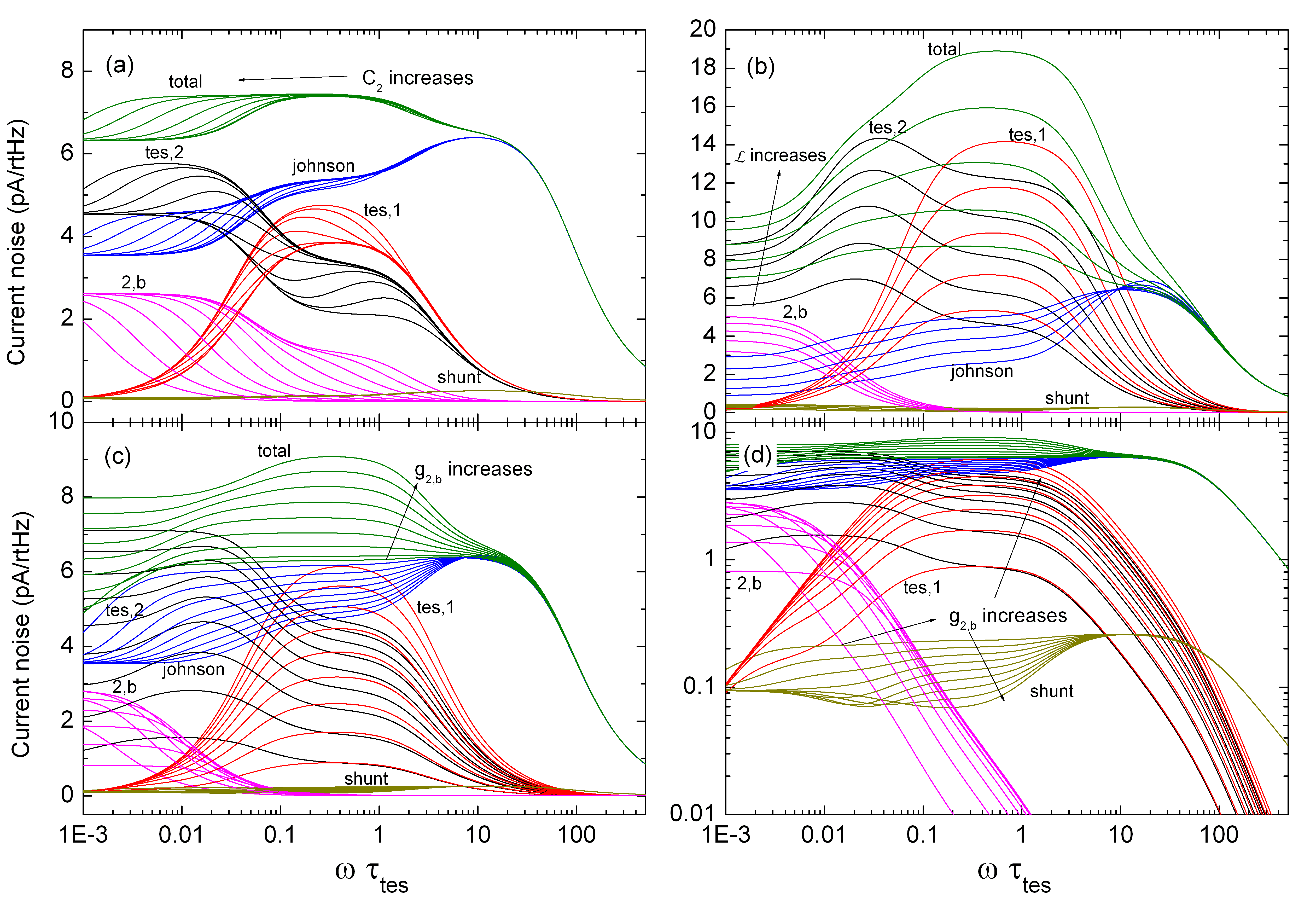}
\caption{(Color online) Current noise of a three-block IH model, (a) with varying $C_{2}/C_{tes}= 0.49, 1.1, 2.48, 5.57, 12.5, 28.2, 63.4, 142.7, 321.1, 722.5$  in linear-log scale.  Increasing $C_{2}$ corresponds to a shift of the onset of the total noise bump to lower frequencies, an addition of a bump feature in $|I_{\omega}|_{tes,2}$ that moves to lower frequencies, a reduction of the $|I_{\omega}|_{tes,1}$ noise component,  and a decrease of the cut-off frequency for the $|I_{\omega}|_{2,b}$ noise component. Mid-frequency Johnson noise also increases. (b) Same, with varying ${\cal L} = 2.48, 3.71, 5.57, 8.35, 12.5$. Increasing ${\cal L}$ corresponds to an increase of the TFN noise levels, and a decrease of the low-frequency Johnson noise. (c) Same, with varying   $c=g_{2,b}/(g_{tes,2}(T_{2})+g_{2,b})= 0.1, 0.2, 0.3, 0.4, 0.5, 0.6,0.7,0.8,0.9$.  Increasing $c$ corresponds to a monotonous increase of the $|I_{\omega}|_{tes,1}$ and $|I_{\omega}|_{tes,2}$ noise components, but a non-monotonous behavior of $|I_{\omega}|_{2,b}$ with maximum amplitude around $c=0.3$, and a shift of the cut-off to higher frequencies.  Johnson noise $|I_{\omega}|_{J}$ decreases in the mid-frequency range. (d) same as (c), but shown in log-log plot. Other parameters in plots (if not varied): ${\cal L}=1.65$, $C_{1}/C_{tes}=10$,  $c=0.5$, $b=g_{tes,1}/(g_{tes,1}+g_{tes,2}(T_{1}))$=0.5, $R_{0}=0.1 \mathrm{\Omega}$, $\beta=1$,  $R_L=0.001 \mathrm{\Omega}$, $I_0=10 \mu$A, $g_{tes,2}(T_{1})=1$ nW/K, $\tau_{el}=L/[R_L+R_0(1+\beta)]=0.015\tau_{tes}$, $\tau_{tes}=C_{tes}/g_{tes,2}(T_{0})$.  }
\label{noiseIH}       % Give a unique label
\end{figure} 

For the IH model, we have plotted a few examples of the noise in Fig. \ref{noiseIH}. It is naturally harder to get as complete picture of the phenomenology as for the two-block models, and here we only discuss an example  where we keep both $g_{tes,1}$ and $g_{tes,2}(T_{0})$ constant, and assume that the thermal exponents for the thermal conductances connected to $C_{2}$ are equal $n=m=4$. In Fig. \ref{noiseIH} (a) we see that an increase of $C_{2}$ corresponds to a shift of the onset of the total noise bump to lower frequencies, but no added bump amplitude in the total noise for this value of ${\cal L}$. An additional bump feature appears in $|I_{\omega}|_{tes,2}$, which moves to lower frequencies with increasing $C_{2}$. In addition, the amplitude of the hanging $|I_{\omega}|_{tes,1}$ noise component is reduced,  and the cut-off frequency for the $|I_{\omega}|_{2,b}$ noise component moves to lower frequencies. Mid-frequency Johnson noise also increases below the frequency of the onset of the $|I_{\omega}|_{tes,1}$ noise component. The effect of the loop gain is seen in Fig. \ref{noiseIH} (b), with the typical result that all the TFN noise components grow and have a higher frequency cut-off with increasing ${\cal L}$, and that the low-frequency Johnson noise is suppressed. Again, the shape of the total noise changes with increasing ${\cal L}$, with $|I_{\omega}|_{tes,1}$ becoming more dominant over $|I_{\omega}|_{tes,2}$. In Fig. \ref{noiseIH} (c), the effect of increasing $g_{2,b}$ is studied, by using the parameter $c=g_{2,b}/(g_{tes,2}(T_{2})+g_{2,b})$. The result is that the total noise level increases and develops a clear bump, whose onset moves to higher frequencies. Both $|I_{\omega}|_{tes,1}$ and $|I_{\omega}|_{tes,2}$ noise components increase monotonously, whereas $|I_{\omega}|_{2,b}$ has a maximum amplitude around $c=0.3$, with its cut-off moving to higher frequencies with $c$. Also, the shape of the $|I_{\omega}|_{tes,2}$ noise changes, with the bump disappearing with high values of $c$. The Johnson noise suppression is strengthened with increasing $c$. The data in Fig. \ref{noiseIH} (c), is also shown in log-log scale in Fig. \ref{noiseIH} (d), highlighting how the shunt noise is suppressed more with increasing $c$, similar to the Johnson noise. A minimum develops, though, for high values of $c$.

More examples of the IH model, and particularily in combination with analysing real impedance and noise data, are discussed in Refs. \cite{kimmo,mikko,kimmothesis}.
   
\section{Conclusions}

We have given here a comprehensive discussion of of all the possible variants of two-block thermal models and their influence on the complex impedance, responsivity and noise of a voltage biased transition edge sensor. The results were derived analytically, and easy-to-use formulas were provided for impedance, responsivity and noise. In addition, results for two variants of simple three-block thermal models were derived, as well. Example plots were generated to show how different parameters of the models affect the observables. In general, a more complex thermal circuit reduces the responsivity and increases the noise at intermediate frequencies, leading thus to performance degradation in terms of the noise equivalent power. The derived theoretical formulas in their current formulation are mainly meant to be used in the analysis of TES experiments, and excellent agreement was already achieved in fitting real TES detector data \cite{mikko,kimmo,kimmothesis}. However, for detector performance optimization one also needs to consider figures of merit such as noise equivalent power, energy resolution, speed and electrothermal stability in more detail. Those performance considerations will be possible in the future for various kinds of TES detectors with more complex thermal circuits, based on the work presented here. 
 
\section*{Acknowledgements}
 This research was supported by Academy of Finland project number 128532 and the Finnish Funding Agency for Technology and Innovation TEKES. We thank K. Kinnunen and M. Palosaari for valuable comments on the manuscript, and L. Gottardi for sharing unpublished work.

\end{document}